\documentclass[twocolumn,superscriptaddress,amsmath,amssymb,aps,prl, sort&compress]{revtex4-1}

\usepackage{times,color,amsthm,graphics,graphicx,bm,bbm,dcolumn}
\usepackage{epsfig}
\usepackage{comment}
\usepackage{graphicx}
\usepackage{xcolor}
\usepackage{booktabs}
\usepackage[colorlinks,urlcolor=black,citecolor=blue,linkcolor=blue]{hyperref}
\usepackage{units}
\usepackage{multibbl}

\begin{document}
%\setpagewiselinenumbers
%\modulolinenumbers[5]
%\linenumbers

%\title{Temperature-driven breakdown of Josephson supercurrent in a unitary superfluid}
%\title{Breakdown of Josephson superflow across the superfluid transition of a unitary Fermi gas}
%\title{Mapping out the conduction properties of a strongly interacting tunnel junction}
%\title{DC response of a strongly interacting tunnel junction across the superfluid transition}
%\title{Anomalous dc response of a strongly interacting tunnel junction across the superfluid transition}. 
%\title{Anomalous tunnel conduction in strongly interacting Fermi gases across the superfluid transition} 
%\title{Anomalous tunneling currents in strongly interacting Fermi gases across the superfluid transition} 
%\title{Anomalous tunneling currents in strongly interacting Fermi gases} 
%\title{Critical breakdown of supercurrents and anomalous tunnel conductance in unitary Fermi gases}
%\title{Supercurrent breakdown and anomalous tunnel conductance in unitary Fermi gases}
%\title{Supercurrent breakdown and anomalous tunneling currents in unitary Fermi gases} 
%\title{Tunneling supercurrents and anomalous normal currents in unitary Fermi gases} 
%\title{Superfluid and anomalous tunneling currents in unitary Fermi gases} 
%\title{Coherent and anomalous tunneling currents in unitary Fermi gases across the superfluid transition}
%\title{Tunneling conduction of unitary fermions across the superfluid transition}
\title{Tunneling transport of unitary fermions across the superfluid transition}

\author{G.~Del~Pace}
\email[] {delpace@lens.unifi.it}
\affiliation{Department of Physics and Astronomy, University of Florence, 50019 Sesto Fiorentino, Italy}
\affiliation{European Laboratory for Nonlinear Spectroscopy (LENS), 50019 Sesto Fiorentino, Italy}
\affiliation{Istituto Nazionale di Ottica del Consiglio Nazionale delle Ricerche (CNR-INO), 50019 Sesto Fiorentino, Italy}
\author{W. J. Kwon}
\affiliation{European Laboratory for Nonlinear Spectroscopy (LENS), 50019 Sesto Fiorentino, Italy}
\affiliation{Istituto Nazionale di Ottica del Consiglio Nazionale delle Ricerche (CNR-INO), 50019 Sesto Fiorentino, Italy}
%\author{M. Inguscio}
%\affiliation{Istituto Nazionale di Ottica del Consiglio Nazionale delle Ricerche (CNR-INO), 50019 Sesto Fiorentino, Italy}
%\affiliation{European Laboratory for Nonlinear Spectroscopy (LENS), 50019 Sesto Fiorentino, Italy}
%\affiliation{Department of Engineering, Campus Bio-Medico University of Rome, 00128 Rome, Italy}
\author{M. Zaccanti} 
\affiliation{European Laboratory for Nonlinear Spectroscopy (LENS), 50019 Sesto Fiorentino, Italy}
\affiliation{Istituto Nazionale di Ottica del Consiglio Nazionale delle Ricerche (CNR-INO), 50019 Sesto Fiorentino, Italy}
\author{G. Roati}
\affiliation{European Laboratory for Nonlinear Spectroscopy (LENS), 50019 Sesto Fiorentino, Italy}
\affiliation{Istituto Nazionale di Ottica del Consiglio Nazionale delle Ricerche (CNR-INO), 50019 Sesto Fiorentino, Italy}
\author{F. Scazza}
\email[] {scazza@lens.unifi.it\\}
\affiliation{European Laboratory for Nonlinear Spectroscopy (LENS), 50019 Sesto Fiorentino, Italy}
\affiliation{Istituto Nazionale di Ottica del Consiglio Nazionale delle Ricerche (CNR-INO), 50019 Sesto Fiorentino, Italy}

\begin{abstract}
We investigate the transport of a Fermi gas with unitarity-limited interactions across the superfluid phase transition, probing its response to a direct current (dc) drive through a tunnel junction. 
As the superfluid critical temperature is crossed from below, we observe the evolution from a highly nonlinear to an Ohmic conduction characteristics, associated with the critical breakdown of the Josephson dc current induced by pair condensate depletion. 
Moreover, we reveal a large and dominant anomalous contribution to resistive currents, which reaches its maximum at the lowest attained temperature, fostered by the tunnel coupling between the condensate and phononic Bogoliubov-Anderson excitations.
Increasing the temperature, while the zeroing of supercurrents marks the transition to the normal phase, the conductance drops considerably but remains much larger than that of a normal, uncorrelated Fermi gas tunneling through the same junction. We attribute such enhanced transport to incoherent tunneling of sound modes, which remain weakly damped in the collisional hydrodynamic fluid of unpaired fermions at unitarity. 
\end{abstract}

\maketitle

%%%%%%%%%
% INTRO %
%%%%%%%%%%

%

Quantum mechanical tunneling underlies many fundamental 
phenomena in physics, 
and it is the backbone for the operation of a variety of electronic devices, ranging from flash memories %to scanning tunneling microscopes. %quantum dots and photovoltaic cells 
to SQUID magnetometers. 
A minimal realization of quantum tunneling is a so-called tunnel junction, created by connecting two conducting materials through a thin insulating layer or potential barrier \cite{Burstein1969}. A tunnel junction represents a unique architecture to understand the %
elementary many-body mechanisms behind mesoscopic transport in quantum systems \cite{Giaever1960, Burstein1969, DattaBook},
%,
%
%
%
hinging essentially on the nature 
of elementary excitations above the ground state \cite{PinesNozieres}. In fermionic systems with attractive interactions, pairing correlations deeply affect both the equilibrium state and its low-energy excitations, %notably 
leading to superfluidity when fermion pairs condense below the critical temperature. 
%In particular, 
The excitation spectrum of %
fermion condensates incorporates both fermionic quasiparticles \cite{Bogoliubov1958, Kivelson1990} and gapless Bogoliubov-Anderson (BA) phonons \cite{Bogoliubov1958, Anderson1958, Combescot2006, ZwergerBook}. While the former correspond to pair-breaking excitations, the latter are 
associated with %
gauge symmetry breaking \cite{Goldstone1961}, and are essential for the condensate to acquire superfluid properties \cite{Landau1941}.

Whereas %electron 
electron transport in superconducting tunnel junctions (STJs) 
is well understood within the Bardeen-Cooper-Schrieffer (BCS) regime of weak attractive interactions, where fermionic degrees of freedom govern both supercurrents and incoherent currents \cite{Blonder1982,Tinkham}, a more intricate interplay between phononic and fermionic excitations is expected to arise in strongly attractive Fermi systems \cite{ZwergerBook,Meier2001,Combescot2006,Uchino2020}.
Atomic Fermi gases near a Feshbach resonance provide a well-controlled % 
framework for addressing two-terminal transport in strongly interacting quantum fluids \cite{Krinner2017}, allowing to reach the universal, unitarity-limited interaction regime \cite{ZwergerBook}. 
In particular, resonant Fermi gases represent
the most robust % 
class of fermionic superfluids and feature 
hydrodynamic behavior even in the normal phase % 
\cite{Hoinka2017,Kuhn2020,Patel2019}. 
This has spurred experimental investigations of mesoscopic transport in unitary superfluid junctions \cite{Stadler2012,Valtolina2015,Husmann2015,Burchianti2018,Husmann2018,Kwon2020,Luick2020}, 
which have revealed 
nonlinear dc current-bias characteristics from multiple Andreev reflections \cite{Husmann2015} or Josephson supercurrents \cite{Kwon2020}. 
%
%%%%%%%%%%%%%%%%%%%%%%%%%%%%%
%%%   RESULTS AND CLAIMS  %%%
%%%%%%%%%%%%%%%%%%%%%%%%%%%%%

In this work, we explore the tunneling conduction of a resonant atomic Fermi gas across the superfluid transition. We show that both supercurrents and normal 
currents react distinctly to the % 
temperature, as they are tied to the amplitude of the superfluid order parameter -- the pair condensate density -- and to its excitation modes. In particular, by measuring the response to a tunable direct current (dc) drive $I_\mathrm{ext}$ in a tunnel junction at varying temperature $T$, we observe the critical breakdown of coherent Josephson transport. 
The dependence of the maximum supercurrent $I_{s,\mathrm{max}}$ on the temperature is % 
captured by a theoretical model %
relying essentially on the thermal depletion of the condensate density, which vanishes at the critical temperature $T_c$. % 
Further, we reveal a large anomalous contribution to the bias-independent conductance, dominating the resistive current branch arising for $|I_\mathrm{ext}| > I_{s,\mathrm{max}}$ at low temperatures, which we ascribe to the conversion of the condensate into phononic BA modes.
Being fueled by the condensate, unlike quasiparticle currents 
in STJs, the ensuing normal current decreases towards $T_c$, above which broken-pair quasiparticles are expected to become the dominant current carriers. Indeed, by measuring the scaling of conductance with the tunneling barrier strength, we distinguish conduction mediated by paired or unpaired fermions. %  
%, 
Above $T_c$ the conductance remains much larger than that measured in a metallic state, even though BA phonons cannot propagate in the absence of the order parameter. We ascribe this to incoherent tunneling of hydrodynamic sound modes, stabilized by elastic collisions between % 
unpaired fermions. Our measurements support scenarios where no significant preformed pair fraction exists at unitarity above $T_c$ \cite{Janko1999, Bergeal2008,Pini2019}.
%
%

%%%%%%%%%%%%%%%%%%%%%%
%%%   EXPERIMENT:  %%%
%%%%%%%%%%%%%%%%%%%%%%
%%%%%%%%%%%%%%%%%%%%%%%%%%%%%%%%%%
%%%  Description  FIGURE 1-2a  %%%
%%%%%%%%%%%%%%%%%%%%%%%%%%%%%%%%%%

\begin{figure}[t]
\centering
\includegraphics[width=72mm]{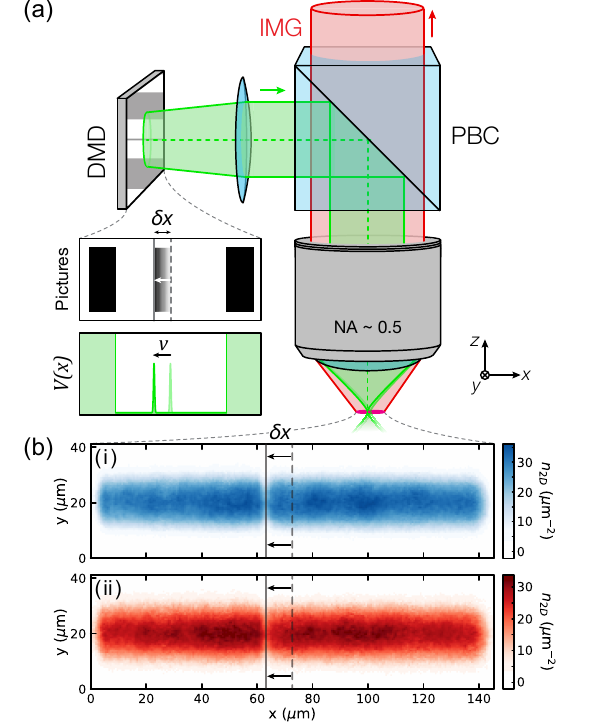}
	\caption{%
	A current-driven tunnel junction between resonant Fermi gases across the superfluid transition. (a) Experimental realization of a tunneling dc drive by dynamical optical potentials. The surface of a DMD displaying the junction geometry, composed of a central barrier and two endcaps, is projected onto the atoms through a $\mathrm{NA}\simeq0.5$ objective, creating a repulsive potential $V(\textbf{r})$. The barrier is set in uniform motion at velocity $v$ by playing a sequence of images on the DMD at the desired framerate, and is swept over a distance $\delta x$ along the $x$-axis.
	(b) In-situ absorption images of the atomic density at unitarity, acquired immediately after completing a $\delta x \simeq 10\,\mu$m barrier translation. The column-integrated density profiles $n_\mathrm{2D}(x,y)$ are shown for an injected current $I_\mathrm{ext} \simeq 3.9 \times 10^5$\,s$^{-1}$ at (i) $T = 0.08(1)\,T_F$ and (ii) $T = 0.18(1)\,T_F$. While no density difference is visible between the two reservoirs for the colder sample (i), the hotter sample (ii) displays a density increase in the left reservoir, producing a chemical potential bias $\Delta\mu \neq 0$ across the junction.
	}
\label{Fig1}
\end{figure}

In our experiment, two tunnel-coupled, strongly interacting atomic reservoirs are created by confining a Feshbach-resonant Fermi gas into a hybrid optical potential, combining a harmonic optical trap with repulsive potentials tailored by a digital micro-mirror device (DMD). %[see Fig.~\ref{Fig1}(a)]. 
Each reservoir contains approximately $N_{R,L} \simeq 4 \times 10^4$ atoms in each of the two lowest hyperfine states of ${}^6$Li. The reservoirs are set initially in thermochemical equilibrium, and their temperature is adjusted  between $T/T_F = 0.07(1)$ and $0.23(1)$,
across the superfluid transition for unitarity-limited interactions at $T_c \simeq 0.21\,T_F$ \cite{Haussmann2008, Pini2019}. Here, $E_F$ and $T_F = E_F/k_B$ are the in-trap % 
Fermi energy and temperature \cite{SM}. 
The optical setup for creating and driving the tunnel junction is illustrated in Fig.~\ref{Fig1}(a). 
The reservoirs are %
separated by a thin, repulsive optical barrier, which is Gaussian along the $x$-direction with a $1/e^2$ width $w_0 \simeq 0.95(9)\,\mu$m and nearly homogeneous along the $y$ and $z$ directions \cite{SM, Kwon2020}. 
Its intensity profile and position are controlled by the DMD, whose surface is imaged onto the atomic sample through a %
high-resolution objective, creating also two sharp axial endcaps. To initialize the junction, the barrier is adiabatically raised at the trap center to the target potential height $V_0$ experienced by one fermion. This creates two identical reservoirs with vanishing relative imbalance $z = (N_L-N_R)/N \simeq 0$, and correspondingly vanishing chemical potential difference $\Delta \mu = \mu_{L} - \mu_{R} \simeq 0$, where $N=N_L+N_R$ and $\mu_{R,L}$ are the fermion chemical potentials in the reservoirs.
% 
%
%
%%%%%%%%%%%%%%%%%%%%
%%%   FIGURE 1   %%%
%%%%%%%%%%%%%%%%%%%%
%%%%%%%%%%%%%%%%%%%%%
%%%%%%%%%%%%%%%%%%%%%%
%
%
To probe the response of the junction to an external dc drive, we set the potential barrier in uniform motion with respect to the gas \cite{Kwon2020,Giovanazzi2000,Levy2007}. % 
The imparted current $I_{\mathrm{ext}}$ is proportional to the barrier velocity $v$, and for a constant total barrier displacement $\delta x\simeq10\,\mu$m, $I_{\mathrm{ext}}=-\bar{z}\,N/2 \times |v|/\delta x$, where $\bar{z} \simeq \pm0.15$ is the relative imbalance at equilibrium for the final barrier position $x=\pm\delta x$.
%
%%%%%%%%%%%%%%%%%%%%%
%
%
By measuring the relative imbalance $z$ after the barrier displacement via in-situ absorption imaging [see Fig.~\ref{Fig1}(b)], we determine the induced potential difference $\Delta\mu=(z-\bar{z})\, E_c\,N/2$. Here, $E_{c} = 2\partial \mu_{L} / \partial N_{L}$ (calculated with $N_L = N/2$) is the effective charging energy of the junction, that is the inverse compressibility of the reservoirs quantifying their density response to a current \cite{Giovanazzi2000,Meier2001}. %

\begin{figure}[t]
\centering
\includegraphics[width=82mm]{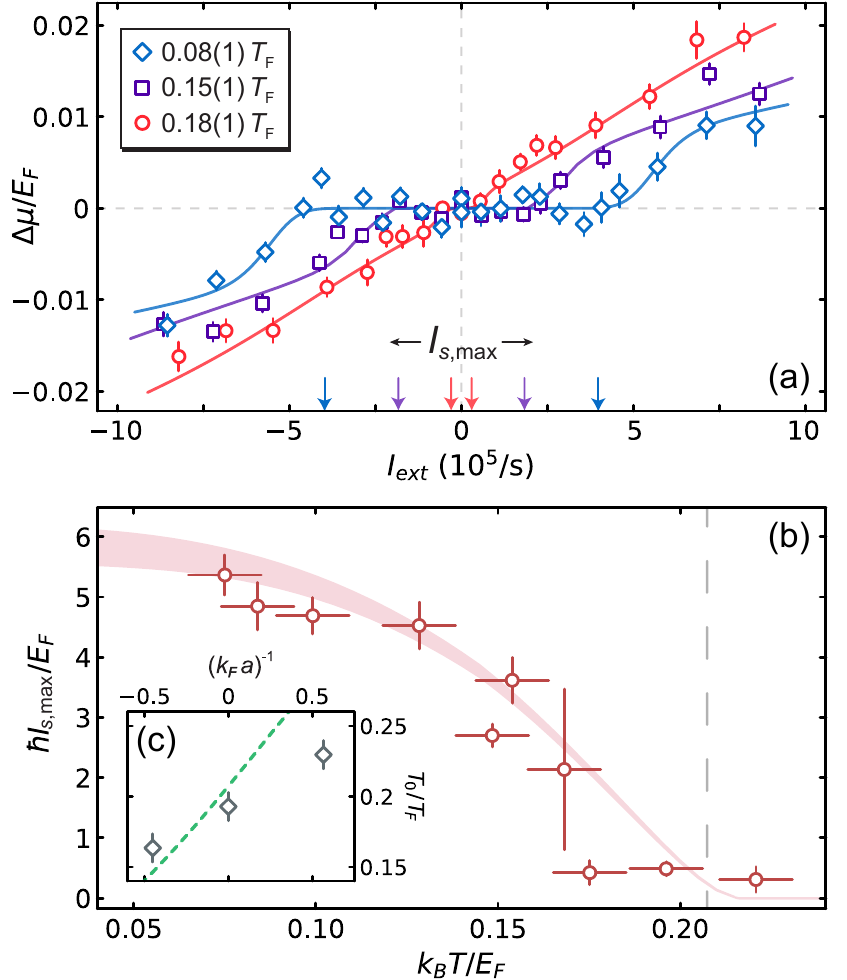}
 \caption{%
 Breakdown of dc tunneling supercurrents in strongly interacting Fermi gases. 
 (a) Current-potential $I-\Delta\mu$ characteristics of the junction at unitarity for different gas temperatures $T/T_F$ (see legend). Solid lines represent fits of experimental data with a RCSJ circuit model \cite{SM}. The fitted values for the maximum supercurrent $I_{s,\mathrm{max}}$ are denoted by arrows on the bottom axis. Error bars denote the s.e.m.~from averaging over $\sim 10$ experimental realizations.
 (b) $I_{s,\mathrm{max}}$ at unitarity as a function of  the reduced temperature $T/T_F$.
 Vertical (horizontal) error bars combine the standard error on $I_{s,\mathrm{max}}$ (temperature) with statistical errors from averaging typically 2 independent extractions. The shaded region indicates the calculated $I_{s,\mathrm{max}}$ (see text), considering a 10\% uncertainty around the nominal barrier width 
 and a 3.5\% uncertainty on the calculated % 
 $E_F$. 
 The barrier height is fixed at $V_0/\mu_0 \simeq 0.7$, where $\mu_0 \approx 0.6\,E_F$ % 
 \cite{Ku2012}. The dashed vertical line marks the predicted critical temperature $T_c$ for the superfluid transition at unitarity \cite{Haussmann2008}.
 (c) Critical temperature $T_0/T_F$ for the disappearance of Josephson currents as a function of interaction strength $(k_F a)^{-1}$. %
 Experimental results are compared with the theoretically obtained $T_c$ from Ref.~\cite{Pini2019} (green dashed line). 
 }
\label{Fig2}
\end{figure}

In a full measurement at fixed temperature, we obtain $\Delta\mu$ as a function of $I_{\mathrm{ext}}$, corresponding to the ``current-voltage'' $I-\Delta\mu$ response of the junction \cite{Kwon2020}. Figure~\ref{Fig2}(a) illustrates the effect of reservoir temperature on such response at unitarity. %, displaying $I-\Delta\mu$ curves obtained at three different temperatures. 
At low temperature, the junction exhibits $\Delta\mu \simeq 0$ for $|I_{\mathrm{ext}}| \leq I_{s,\mathrm{max}}$, implying vanishing resistance below a maximum current $I_{s,\mathrm{max}}$. By raising the temperature, smaller values of $I_{\mathrm{ext}}$ suffice to yield a finite $|\Delta\mu| > 0$. Eventually, $I_{s,\mathrm{max}}$ vanishes and any non-zero applied current produces a chemical potential difference, hence the junction becomes fully resistive. % 
Crucially, the junction crosses over from a nonlinear to an Ohmic current-potential characteristic, reflecting the phase transition of the system from the superfluid to the normal state \cite{Anderson1969, Tinkham, BaroneBook}.
Within the linear response regime $\Delta\mu \ll \mu_0$ explored here, where $\mu_0$ is the peak chemical potential in the absence of barrier, the current injected through the junction can be decomposed as \cite{BaroneBook, Tinkham}:
\begin{equation}
\begin{aligned}
&I_{\mathrm{ext}} = I_{s}(\varphi) + G \Delta\mu + C \Delta\dot{\mu}\,,
\end{aligned}
\label{eq:totalcurrent}
\end{equation}
where $\varphi = \varphi_L - \varphi_R$ is the phase difference between the pair condensates in the reservoirs, $I_{s}(\varphi)$ is the current-phase relation of the junction, $G$ is the bias-independent tunneling conductance, and $C = 1/E_c$ is the effective junction capacitance. For barrier heights $V_0/\mu_0\sim 1$, corresponding to small junction transmissions $|t_p|^2 \ll 1$, where $t_p$ is the tunneling amplitude of a single pair, one can retain only the first two orders in the tunnel coupling \cite{Meier2001,Kwon2020} such that $I_{s}(\varphi) \simeq I_1 \sin \varphi + I_2 \sin 2\varphi$, where the coefficients $I_n$ are of order $|t_p|^n$ \cite{Bloch1970,Meier2001}. In the absence of an initial bias potential, 
$I_{s,\mathrm{max}} = \max_{\varphi \in [0,\pi)} |I_s(\varphi)|$ sets the largest value of $I_\mathrm{ext}$ for which the junction exhibits zero normal current, i.e., $\Delta\mu = 0$ for $|I_\mathrm{ext}| \leq I_{s,\mathrm{max}}$. 
The observed zero-potential plateaus [see Fig.~\ref{Fig2}(a)] 
gauge therefore the maximum Josephson supercurrent flowing through the junction \cite{Kwon2020}, which vanishes for $T\geq T_c$ in the normal state. 
On the other hand, %
the charging rate $G E_c$ sets the timescale for the junction resistive response.
Combining Eq.~\eqref{eq:totalcurrent} with the Josephson-Anderson relation $\hbar\dot{\varphi} = - \Delta \mu$ yields the resistively and capacitively shunted junction (RCSJ) circuit model, widely applied to STJs \cite{BaroneBook}. To characterize the junction response, we fit the measured $I-\Delta\mu$ curves with the numerical solution of such RCSJ model \cite{SM}, extracting $I_{s,\mathrm{max}}$ and $G$ for each temperature. 

%%%%%%%%%%%%%%%%%%%%
%%%   FIGURE 2   %%%
%%%%%%%%%%%%%%%%%%%%
%%%%%%%%%%%%%%%%%%%%%%%%%%%%%%%%%%%%%%%%%%
%%%%%%%%%%%%%%%%%%%%%%%%%%%%%%%%%%%%%%%%%%

%%%%%%%%%%%%%%%%%%%%%%%%%%%%%%%%%%
%%%    Description FIGURE 2b    %%%
%%%%%%%%%%%%%%%%%%%%%%%%%%%%%%%%%%

%
Figure\,\ref{Fig2}(b) shows the obtained $I_{s,\mathrm{max}}$ at unitarity as a function of the reservoir temperature. 
The observed trend %
resembles that predicted within BCS theory by the Ambegaokar-Baratoff (AB) formula, $I_{s,\mathrm{max}} = \pi\Delta/2 \times G_n\,\text{tanh}(\Delta/2k_BT)$ \cite{Ambegaokar1963b}, where $\Delta$ is the superconducting gap, and $G_n$ is the conductance of the junction in the normal state right above $T_c$, % 
set essentially by the single-fermion tunneling probability. 
However, %this 
the weak-coupling AB result is not expected 
to hold at strong interactions, even for $T=0$ \cite{Spuntarelli2007, Zaccanti2019}. Therefore, we model our system by generalizing to finite temperatures the effective theory presented in Refs.~\cite{Zaccanti2019,Kwon2020}, where the Josephson supercurrent is explicitly linked to the condensate density, obtaining the shaded curve in Fig.~\ref{Fig2}(b) \cite{SM}. For this, we exploit %
Luttinger-Ward calculations of the condensate fraction for a homogeneous unitary Fermi gas \cite{Haussmann2007}, as well as its equation of state \cite{Ku2012} in the local density approximation (LDA). 
With no free parameters, the model %
quantitatively reproduces the experimental data, evidencing a firm connection between $I_{s,\mathrm{max}}$ and the condensate density at any temperature. Yet, some discrepancy appears when approaching $T_c$. While the calculated $I_{s,\mathrm{max}}$ vanishes essentially at $T_c$ [dashed vertical line Fig.~\ref{Fig2}(b)], the measured one sharply drops to nearly zero already at $T_0<T_c$.
Such deviation could result from the radial inhomogeneity of the gas, causing different shells to undergo the superfluid transition at different $T$. % 
Furthermore, a finite $\Delta\mu$ at $T<T_c$ could develop from thermal fluctuations in the superfluid state \cite{Anderson1969,Halperin2011}. In particular, stochastic thermal phase slips, whose probability scales as $\exp(-2\hbar I_{s,\mathrm{max}}/k_B T)$ \cite{Halperin2011}, are expected to become relevant when $\hbar I_{s,\mathrm{max}} \lesssim E_F$,
which in our system occurs at $T \gtrsim 0.17\,T_F$. %
We determine the critical temperature $T_0$ for the breakdown of Josephson supercurrents through a piecewise function fit of $I_{s,\mathrm{max}}(T)$ \cite{SM}. At unitarity $T_0 = 0.19(1)\,T_F$, consistent with what expected from thermal phase slips. The same procedure is used to determine $T_0$ at different interaction strengths, parametrized by $(k_Fa)^{-1}$ where $k_F = \sqrt{2mE_F}/\hbar$ and $a$ is the $s$-wave scattering length, as displayed in Fig.~\ref{Fig2}(c). The observed monotonic increase of $T_0$ from 
negative to positive couplings reflects that of the superfluid
critical temperature $T_c$ \cite{ZwergerBook,Pini2019}. %

%

%%%%%%%%%%%%%%%%%%%%%%%
%%%   CONDUCTANCE   %%%
%%%%%%%%%%%%%%%%%%%%%%%

\begin{figure}[t]
\centering
\includegraphics[width=82mm]{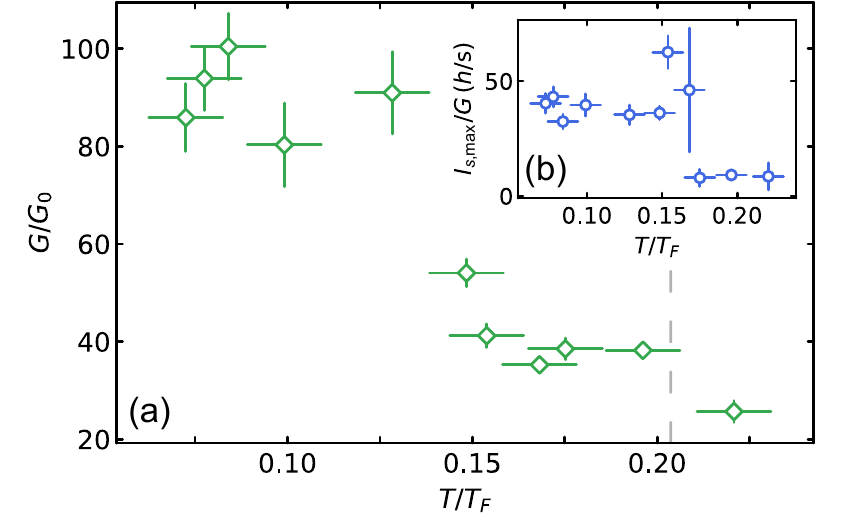}
 \caption{Bias-independent tunneling conductance $G$ as a function of $T/T_F$, measured at unitarity under the same experimental conditions of Fig.~\ref{Fig2}. %
 (a) $G$ is normalized to the conductance per spin component $G_0= 102(15)\,h^{-1}$ of a non-interacting Fermi gas, measured at $T/T_F = 0.21(1)$ with an equivalent barrier height $V_0/\mu_0 \simeq 0.73$. Vertical (horizontal) error bars combine the standard error on the extracted conductance (temperature) with statistical errors from averaging typically $2$ independent extractions. 
 The dashed vertical line locates the theoretical critical temperature $T_c$ %
 at unitarity \cite{Haussmann2008}.
 (b) Experimental ratio 
 $I_{s,\mathrm{max}}/G$ as a function of $T/T_F$. % 
 }
\label{Fig3}
\end{figure}

The bias-independent tunneling conductance $G$ obtained at unitarity is shown in Fig.~\ref{Fig3}(a) as a function of temperature, normalized to the measured (normal-state) conductance $G_0 = 102(15)\,h^{-1}$ of a non-interacting Fermi gas at $V_0 \simeq 0.7\,E_F$. 
$G$ %
increases monotonically with decreasing $T$, taking values as large as $G\!\sim\!10^2\,G_0\!\sim\! 10^4\,h^{-1}$, and greatly exceeds $G_0$ for any $T$. We compute the normal-state conductance of the junction considering zero-temperature ideal fermionic reservoirs within LDA \cite{SM}, obtaining $G_n = 160(29)\,h^{-1}$, in reasonable agreement with the measured $G_0$. 
The mismatch between the measured $G$ and $G_{0}$ (or the estimated $G_n$) indicates that %, unlike in STJs, 
normal currents in our neutral gas junction does not arise from incoherent pair or quasiparticle tunneling, but rather from collective bosonic excitations \cite{Burchianti2018, Uchino2020}. 
In the small bias regime at low temperature $\Delta\mu \ll \Delta$, broken pairs are energetically suppressed as $\exp(-\Delta/k_BT)$, and the only accessible excitation out of the condensate are gapless low-momentum BA phonons \cite{Combescot2006, Hoinka2017}. Only at %
$T \sim T_c$, fermionic quasiparticles proliferate, and their incoherent tunneling is expected to take over as main conduction mechanism.
Conversely, pair-breaking processes in STJs typically determine both the maximum Josephson supercurrent and normal tunneling current \cite{Tinkham},
while no key role is played by BA phonons that are lifted into gapped plasmons by the Coulomb repulsion \cite{Anderson1958}.
%

%%%%%%%%%%%%%%%%%%%%
%%%   FIGURE 3   %%%
%%%%%%%%%%%%%%%%%%%%
%%%%%%%%%%%%%%%%%%%%%%%%%%%%%%%%%%%%%%%%%%
%%%%%%%%%%%%%%%%%%%%%%%%%%%%%%%%%%%%%%%%%%

A phononic contribution to the dc conductance, arising from the tunnel coupling between the condensate in one reservoir and sound modes in the other reservoir, has indeed been predicted both for weakly interacting BECs \cite{Meier2001} and, very recently, for %
neutral fermionic superfluids in the BCS regime \cite{Uchino2020}. %
Such \emph{anomalous} contribution remains finite at $T=0$, being fostered by the condensate and its gapless phononic excitations. 
Additionally, a \emph{normal} finite-$T$ contribution is expected to 
grow as $\sim T^4$ as a consequence of incoherent tunneling of thermally populated phonons \cite{ZwergerBook,Meier2001}. While a full description of tunneling transport in the unitary gas remains an open theoretical challenge, %
fermionic quasiparticle transport appears essentially irrelevant with respect to the experimentally obtained conductance, %
providing thus strong evidence for anomalous tunneling currents in unitary superfluids. 
The anomalous character of low-temperature conductance is further confirmed by considering the experimental ratio $I_{s,\mathrm{max}}/G$, displayed in Fig.~\ref{Fig3}(b). 
Whereas in STJs this quantifies the superconducting gap via the AB formula \cite{Ambegaokar1963b,Tinkham}, the approximately constant value observed here until nearby $T_0$ demonstrates the relation between anomalous conductance and condensate density. %  
Upon increasing the temperature towards $T_c$, the anomalous term is expected to extinguish its effect, and significant incoherent tunneling should set in. However, in contrast to 
collisionless gases, our strongly interacting fluid supports weakly damped hydrodynamic sound even in the normal phase above $T_c$ \cite{Kuhn2020,Patel2019}, with no distinct signature emerging % 
across the superfluid transition. 
%\footnote{The collisional hydrodynamic character of the system is fundamental to also ensure near-instantaneous equilibration of the reservoirs while the current drive is active \cite{Krinner2017}.}.
Therefore, %
phonon tunneling is not expected to fade at $T_c$, explaining the observed trend of $G$: despite dropping by nearly an order of magnitude from its maximum, $G$ remains much larger than $G_0$ also for $T \gtrsim T_c$.

%

%%%%%%%%%%%%%%%%%%%%
%%%   FIGURE 4   %%%
%%%%%%%%%%%%%%%%%%%%
%%%%%%%%%%%%%%%%%%%%%%%%%%%%%%%%%%%%%%%%%%
\begin{figure}[b]
\centering
\vspace*{-10pt}
\includegraphics[width=86mm]{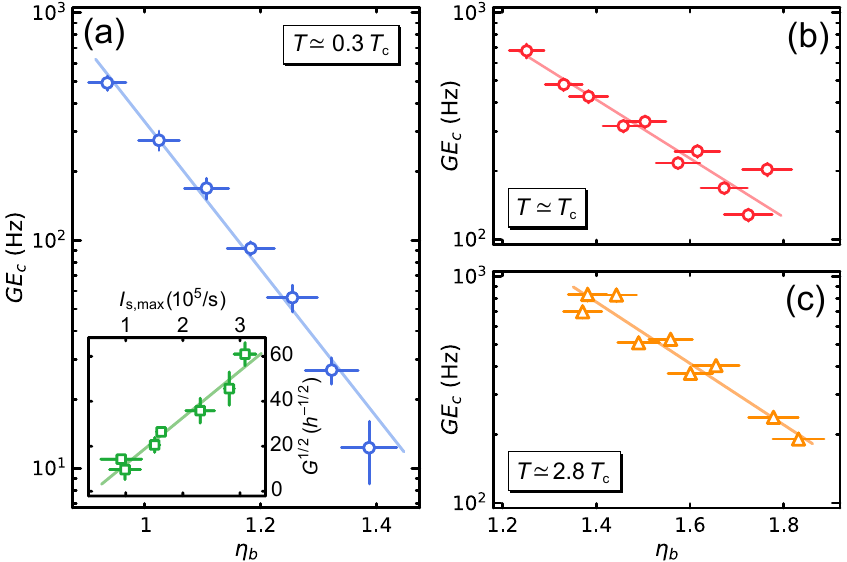}
 \caption{
 Charging rate $G E_c$ as a function of the adimensional barrier strength $\eta_b = (k_F d) \times \sqrt{V_0/E_F}$ \cite{SM} %
 for (a) a superfluid unitary gas at $T=0.06(1)\,T_F \simeq 0.3\,T_c$, (b) a unitary gas at $T = 0.20(1)\,T_F \simeq T_c$, and (c) a normal attractive Fermi gas with $(k_F a)^{-1} \simeq -0.82$ at $T = 0.20(1)\,T_F \simeq 2.8\,T_c$ \cite{gor1961}. In panel (a), $E_F/h \simeq 6$\,kHz (data taken from Ref.~\cite{Kwon2020}), while $E_F/h \simeq 11$\,kHz in panel (b) and (c). Vertical and horizontal error bars result from the standard error on $G$ and the experimental uncertainty on $V_0$, respectively. Solid lines represent linear fits of $\text{log}\,GE_c$.  In the inset of panel (a), the measured $G^{1/2}$ and $I_{s,\mathrm{max}}$ are compared within the same $\eta_b$ range. %
 }
 \vspace{10pt}
\label{Fig4}
\end{figure}
%%%%%%%%%%%%%%%%%%%%%%%%%%%%%%%%%%%%%%%%%%

%%%%%%%%%%%%%%%%%%%%%%%%%%%%%%%%%%
%%%    Description FIGURE 4    %%%
%%%%%%%%%%%%%%%%%%%%%%%%%%%%%%%%%%

The behavior of tunnel conductance provides also information about the nature of current carriers, being sensitive to whether transport is mediated by pairs or unpaired fermions. 
In absence of pair breaking, resistive currents arise only at second order in the tunnel coupling between pairs in the reservoirs \cite{Meier2001,Uchino2019}, namely $G \propto |t_p|^2$. We directly confirm this scaling by comparing the conductance and the supercurrent amplitude at varying barrier strength, finding $G^{1/2} \propto I_{s,\mathrm{max}} \sim |t_p|$ at low temperature [inset of Fig.~\ref{Fig4}(a)]. On the other hand, when broken pairs prevail, $G \propto |t_F|^2$ is expected to reflect the single-fermion tunneling probability $|t_F|^2$. %, that is $G \propto |t_F|^2$.
In the tunneling limit, both $|t_{p}|$ and $|t_{F}|$ decrease exponentially with the adimensional barrier strength $\eta_b = k_{F} d\,\sqrt{V_0/E_F}$, $d = 0.6\,w_0$ being the effective barrier size \cite{SM}. Yet, it can be shown that $\log |t_{p}| \approx 2 \log |t_{F}|$, when accounting for the different mass and polarizability of bound pairs and unpaired atoms \cite{SM}.
Figure~\ref{Fig4}(a)-(c) compare the measured scaling of the charging rate $G E_c$ with $\eta_b$,
for three distinct regimes across the superfluid transition. In all regimes, we observe 
that $\log GE_c \propto -\eta_b$, as expected. Remarkably, a unitary gas at $T \simeq T_c$ [panel (b)] exhibits a weaker dependence of $GE_c$ on $\eta_b$ than a superfluid unitary gas $T \simeq 0.3\,T_c$ [panel (a)], while its behavior is instead equivalent to that of a normal attractive Fermi gas at $T > T_c$ [panel (c)]. Performing linear fits of $\log G E_c$, we find a ratio of 2.5(3) between the slopes in the superfluid and critical unitary regimes, nearly compatible with the factor of $2$ expected between tightly-bound pairs and free fermions. 
Such observations point to a change 
of resistive current carriers from pairs to single fermionic quasiparticles, upon crossing $T_c$ from below, with no analogue in electron systems. Further, our measurements suggest that for $T= 0.20(1)\,T_F$ incoherent transport at unitarity is dominated by unpaired fermions, although weak ``pseudogap'' correlations \cite{Haussmann2009,Magierski2009,Pini2019,Jensen2019}, to which low-momentum phonons are essentially insensitive \cite{Patel2019}, may exist -- and could be probed through spin conductance \cite{Sekino2020}.

%

%%%%%%%%%%%%%%%%%%%%%%%%%
%%%    CONCLUSIONS    %%%
%%%%%%%%%%%%%%%%%%%%%%%%%

%
In conclusion, we have %
demonstrated that tunneling currents constitute a powerful probe of the superfluid order parameter of unitary fermions at finite temperatures, providing a striking signature of the superfluid phase transition \cite{Riedl2011,Ku2012,Sidorenkov2013}.
The order parameter is found to impact both zero-resistance and Ohmic conduction, feeding Josephson and anomalous normal currents, hence presenting significant differences between transport in neutral and charged quantum fluids. 
We anticipate the observed
anomalous low-resistance conduction to be a generic feature, not restricted to the large-area, multimode junction used here, as suggested by recent theoretical investigations of weakly interacting neutral superfluids at quantum point contacts \cite{Uchino2019,Uchino2020}.
In the future, it will be interesting to focus on the regime $T \simeq T_c$, where thermally activated phase slips modify the current-potential characteristics \cite{Anderson1969,Halperin2011} and conduction may be influenced by pairing fluctuations \cite{Scalapino1970, Anderson1970, Bergeal2008}. % 
The anomalous and normal contributions to Ohmic currents could be disentangled by measuring heat transport \cite{Husmann2018,Uchino2019,Uchino2020}.
Finally, % 
realizing a spin-current drive within the same 
tunneling geometry
will enable to probe spin correlations in the pseudogap regime \cite{Sekino2020}.

\vspace*{0pt}
\begin{acknowledgements}
We are indebted to Eugene Demler, Tilman Enss, Massimo Inguscio, Verdiana Piselli, Carlos Sa De Melo, Giancarlo Strinati, and Wilhelm Zwerger for inspiring discussions. We thank Bernhard Frank for providing us with finite-temperature Luttinger-Ward calculations of the condensed fraction of the homogeneous unitary Fermi gas, Michele Pini for providing us with self-consistent calculations of the critical temperature, and Riccardo Panza for assistance during the construction of the experimental setup. This work was supported by the European Research Council under GA no.~307032, 
Fondazione Cassa di Risparmio di Firenze project QuSim2D 2016.0770, 
the Italian MIUR under the PRIN2017 project CEnTraL, and EU's Horizon 2020 research and innovation programme under the Qombs project FET Flagship on Quantum Technologies GA no.~820419, and Marie Sk\l{}odowska-Curie GAs no.~705269 and no.~843303. %
\end{acknowledgements}

\vspace*{-5pt}

%%%% SUPPLEMENTARY %%%
\newpage

\setlength{\belowcaptionskip}{-5pt}
\renewcommand{\baselinestretch}{1.25}

\newcommand{\bra}[1]{\mbox{\ensuremath{\langle #1 \vert}}}
\newcommand{\ket}[1]{\mbox{\ensuremath{\vert #1 \rangle}}}
\newcommand{\Li}{$^{6}$Li }
\hyphenation{Fesh-bach}

\newcommand{\bcirc}{\textcolor{blue}{$\bigcirc$}}
\newcommand{\beq}{\begin{equation}}
\newcommand{\eeq}{\end{equation}}

\renewcommand{\thefigure}{S\arabic{figure}}
\setcounter{figure}{0}
\renewcommand{\theequation}{S.\arabic{equation}}
\setcounter{equation}{0}
\renewcommand{\thesection}{S.\arabic{section}}
\setcounter{section}{0}
\renewcommand{\thetable}{S\arabic{table}}
\setcounter{table}{0}

\renewcommand{\theHequation}{Supplement.\theequation}
\renewcommand{\theHfigure}{Supplement.\thefigure}

\setlength{\tabcolsep}{18pt}

\onecolumngrid

%\newpage

\setcounter{equation}{0}
\setcounter{figure}{0}
\setcounter{table}{0}

%%%%%%%%%%%%%%%%%%%%%%%%%%%%%%%%%%%%%%%%%%%%%%%%%%%%
%\clearpage

\newpage
\begin{center}
\textbf{\large Supplemental Material\\[4mm] 
\Large Tunnel conduction of unitary fermions across the superfluid transition}\\[4mm]
G. Del Pace,$^{1,2,3,\ast}$
W. J. Kwon,$^{2,3}$
M. Zaccanti,$^{2,3}$
G. Roati,$^{2,3}$ and
F. Scazza$^{2,3,\dagger}$
\\[2mm]
\emph{\small $^1$ Department of Physics and Astronomy, University of Florence, 50019 Sesto Fiorentino, Italy}\\
\emph{\small $^2$ Istituto Nazionale di Ottica (CNR-INO), 50019 Sesto Fiorentino, Italy}\\
\emph{\small $^3$ European Laboratory for Nonlinear Spectroscopy (LENS), 50019 Sesto Fiorentino, Italy}
\end{center}
\vspace*{-10pt}
\begin{center}
\emph{\small ${}^\ast$ E-mail: delpace@lens.unifi.it}\\
\emph{\small ${}^\dagger$ E-mail: scazza@lens.unifi.it}
\end{center}

\normalsize 

\setcounter{page}{1}

\section{Experimental sample preparation}\label{Sec:SamplePrep}

To realize two tunnel-coupled fermionic reservoirs, we first produce a degenerate unitary Fermi gas confined within an optical dipole trap. In particular, we evaporate a balanced mixture of the two lowest hyperfine $\ket{F = 1/2, m_F = \pm 1/2}$ states of $^6$Li, labeled as $\ket{1}$ and $\ket{2}$, in a cigar-shaped crossed optical dipole trap on top of their broad scattering resonance at 832\,G \cite{Valtolina2015sm, Burchianti2018sm}. After completing the evaporation stage, we re-compress the crossed dipole trap by increasing the power of both trapping lasers, ensuring a sufficient depth to hold large atomic samples even at the highest target temperature $T/T_F \simeq 0.23$. The final trapping configuration is identical for all measurements presented in Figs.~1--3 of the main text, and it is characterized by the harmonic frequencies $(\omega_x, \omega_y, \omega_z) \simeq 2\pi \times (17, 300, 290)$\,Hz (measured at unitarity). 
We precisely adjust the temperature of the cloud within the range $T/T_F = 0.07 - 0.23$ through controlled parametric heating. Here, $T_F = E_F/k_B$ and $E_F = \hbar \bar{\omega} (6 N_\mathrm{tr})^{1/3}$ is the harmonic-trap Fermi energy, with $\bar{\omega} = (\omega_x \omega_y \omega_z)^{1/3}$ the mean trap frequency and $N_\mathrm{tr}$ the total number of atoms per spin state in the trap.
For this, we modulate the power of one of the trapping beams at about twice the frequency of the radial harmonic trapping during the evaporation stage, setting the modulation amplitude and the heating pulse duration to produce the desired target temperature. Additionally, we adapt the minimum trap depth before the re-compression so as to keep the atom number fixed to $N_\mathrm{tr} \simeq 1.5 \times 10^5$ per spin state at any final temperature. 
After an equilibrated unitary sample has been obtained, to perform measurements away from unitarity we adiabatically sweep the magnetic field to the desired target value between $B \simeq 795$\,G and $B\simeq 920$\,G to tune the inter-atomic $s$-wave scattering length $a$ in 50\,ms, obtaining samples with $1/k_Fa$ between $+0.53$ and $-0.82$. 
The magnetic contribution to the harmonic confinement, resulting from the curvature of the field created by the Feshbach coils, introduces a small variation of the axial trap frequency of about 5\% in the interval between $B \simeq 795$\,G and $B\simeq 920$\,G.

To create the atomic tunnel junction, we exploit a hybrid attractive and repulsive optical trap. This consists of a repulsive optical potential, controlled by a digital micromirror device (DMD) illuminated with $532$\,nm blue-detuned light, and the red-detuned crossed dipole trap, where the degenerate sample is initially produced as described above. In particular, we project onto the harmonically trapped atomic cloud a compound image of a barrier and two endcaps, which realizes the junction geometry shown in Fig.~1(b). The barrier is shone along the $z$-direction and it is homogeneous along the $y$-direction, while it has a Gaussian profile with $1/e^2$ width $w_0 = 0.95\,\mu$m along the axial $x$-direction, as characterized in Ref.~\cite{Kwon2019}.
To initialize the barrier at the trap center, we finely adjust the position of the crossed dipole trap by controlling the position of one of its two laser beams with an acousto-optical deflector. 
The potential height $V_0$ of the barrier is calibrated by shining a DMD-generated homogeneous square-shaped repulsive obstacle of variable optical intensity onto a weakly interacting BEC ($1/k_F a \simeq 9$). By measuring the number of atoms contained in such square potential region as a function of the power illuminating the DMD, we calibrate the intensity at which $V_0$ matches the known chemical potential of the gas. For this purpose, we exploit the known semi-ideal density equation of state for a three-dimensional Bose gas (see Ref.~\cite{Kwon2019} for additional details).

To realize the external dc current drive through the junction, the barrier is set in uniform motion with respect to the sample by displaying a sequence of pictures on the surface of the DMD at the desired frame rate. A corresponding sequence of equidistant trigger pulses is sent to the DMD to control the switching from the displayed picture to the following one, within a pre-loaded sequence of pictures where the barrier is displaced at increments of one DMD-pixel (see Ref.~\cite{Kwon2019} for further details about the sequence creation). 
By varying the time interval between the triggers, we adjust the velocity of the barrier, i.e., the current injected into the junction. %

\section{Temperature extraction}\label{Sec:SM_UFGthermo}
%Thermometry

%
The temperature of the atomic sample is extracted by analyzing high-resolution images of the in-situ atomic density, acquired just before the DMD-generated repulsive potentials are turned on. 
For this purpose, we employ high-intensity absorption imaging through the same high-resolution objective displayed in Fig.~1(a), applying a short pulse of $4 \, \mu$s with an intensity of about $3$ times the saturation intensity of the imaging transition, thereby limiting undesired blurring from spontaneous emission.
The sample temperature is obtained by analyzing the mean density profile of at least $10$ experimental realizations for each measurement. A typical averaged image is shown in Fig.~\ref{FigS1}(a), folded in half by further averaging along the axial $x$-direction.
We have checked that by ramping the repulsive potentials on and off, the sample temperature remains unaffected, verifying the adiabaticity of our loading procedure into the two-terminal junction geometry. Therefore, we conclude that the temperature extracted from the harmonically trapped cloud represents a reliable thermometer of the system confined into the final junction potential.

\subsection{Two-dimensional non-interacting fit}

\begin{figure}[t]
\centering
\includegraphics[width=86mm]{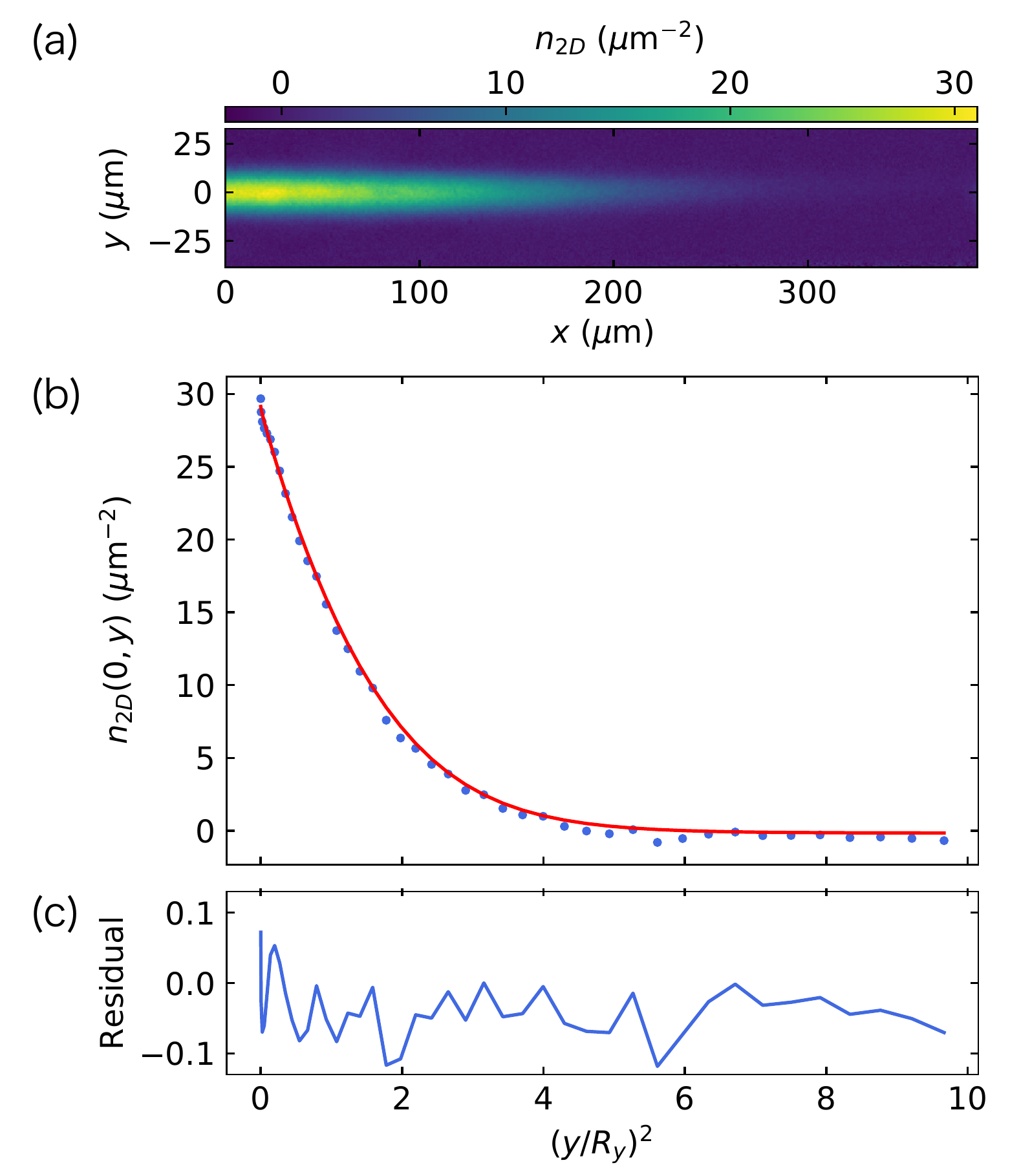}
 \vspace*{-10pt}
 \caption{Temperature extraction through a two-dimensional non-interacting Fermi gas fit. (a) In-situ image of a unitary Fermi gas ($1/k_Fa \sim 0$) in the crossing optical dipole trap at $T/T_F = 0.18(1)$. % 
 The image consist of an average over about $10$ experimental realizations, that for each temperature is used for thermometry, folded in half cloud by averaging the two axial sides.
 (b)-(c) Comparison of data with the non-interacting Fermi gas fit.
 In panel (b), a radial cut of the two-dimensional density profile (blue dots) is compared with the fit (red line). The $y$ axis is presented in adimensional unit, where $R_y = 8.05(1) \, \mu$m is the radial cloud size extracted from the fit.
 In panel (c), the relative residuals of the fit, defined as the difference between the data and the fitted function re-scaled by the average density, are plotted using the same adimensional units.
 }
\label{FigS1}
\end{figure}

As a first estimation of the unitary cloud temperature, we perform a phenomenological fit of the acquired two-dimensional density profile with that of a harmonically trapped non-interacting Fermi gas \cite{Varenna2008}:

\begin{equation}
n_{2D}(x,y) = n_{2D,0} \,\, {\mathrm{Li}_{2}\!\left(-\exp{\left[\beta\mu_0 - \left( {x^2/R_x^2}  + {y^2/R_y^2} \right)\, f\!\left(e^{\beta\mu_0}\right) \right]}  \right) \over \mathrm{Li}_{2}\!\left(-\exp{\left(\beta\mu_0\right)}\right)}
\bigskip
\end{equation}
where $\beta = 1/k_B T$, $\mu_0$ is the chemical potential, $R_j$ is the effective cloud size along the $j$-axis, and $f(x) = ((1+x)/x) \, \ln(1+x)$. 
As already found in previous studies \cite{Varenna2008}, such profile is capable of excellently reproducing the measured density profiles [see Fig.~\ref{FigS1}(b)-(c)], despite not accounting for the exact equation of state of the unitary gas. The effective degeneracy parameter $\tilde{T}/T_F$ is extracted from the fitted $\beta\mu_0$ as $\tilde{T}/T_F = \left[-6\, \mathrm{Li}_{3}\left( - e^{\beta\mu_0} \right)\right]^{-1/3}$. Given that the variation of $\mu_0$ with $T$ is small in the experimentally explored range, the actual degeneracy parameter $T/T_F$ of the cloud is well approximated by $T/T_F \simeq \sqrt{\xi} \times \tilde{T}/T_F$ \cite{Varenna2008}, where $\xi \simeq 0.37$ is the universal Bertsch parameter \cite{Ku2012sm}.
This method is particularly robust against fluctuations of the image background, since it relies on a two-dimensional fitting procedure, and yields thus a reliable estimation of the cloud temperature. To back up this estimation, we have implemented two additional methods for thermometry, based on the known equation of state of the unitary Fermi gas, that are presented in the following Section.

\subsection{Thermometry via the unitary Fermi gas equation of state}

\begin{figure}[t]
\centering
\includegraphics[width=160mm]{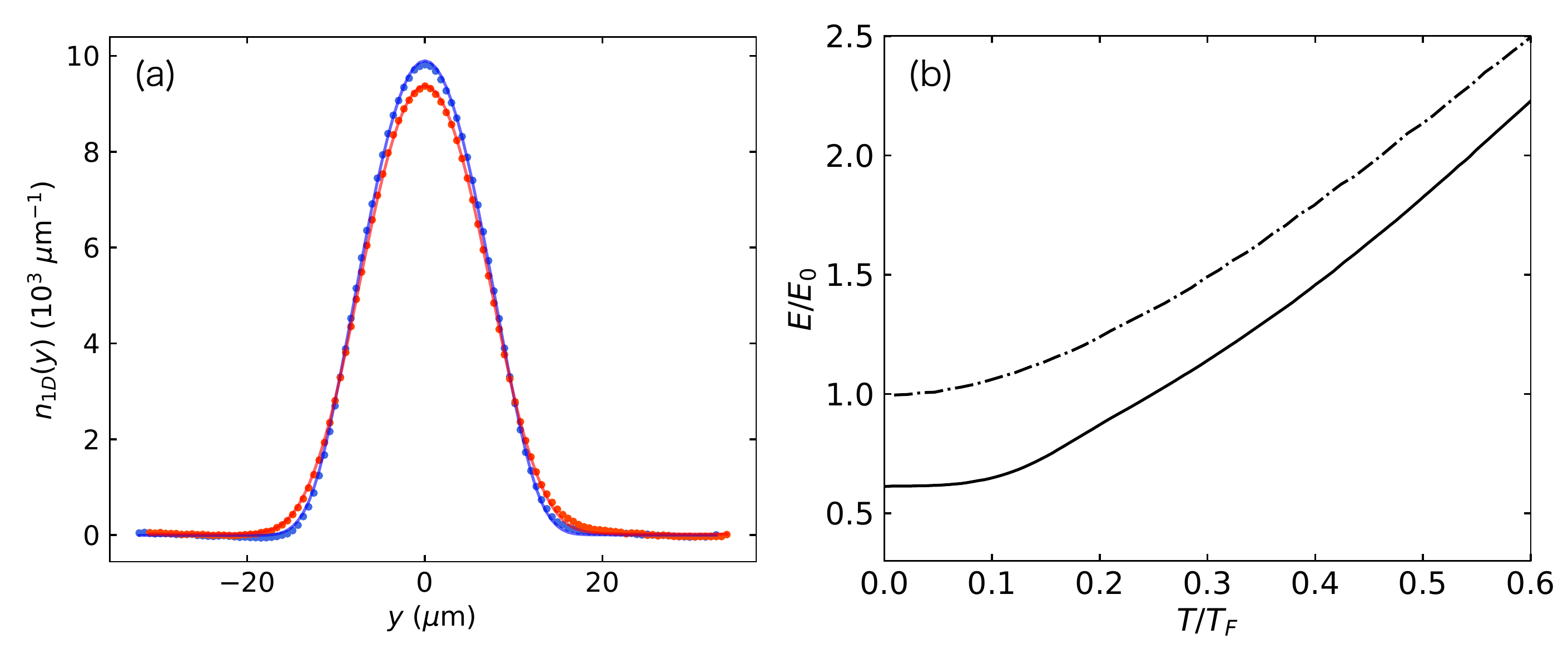}
 \caption{Temperature extraction through the unitary Fermi gas EoS.
 (a) Experimental one-dimensional density profiles and corresponding fits with Eq.~\eqref{Eq:fit1D_eos} at two different temperatures. The fits yield $T/T_F = 0.077(5)$ and $T/T_F = 0.125(3)$ for the blue and the red profile, respectively.
 (b) Total energy $E$ as a function of degeneracy parameter $T/T_F$ for unitary (solid line) and non-interacting (dot-dashed line) Fermi gases \cite{guajardo2013, husmann2015}. In both cases, $E$ is normalized to the total energy $E_0 =  \frac{3}{4} N E_F$ of a zero-temperature non-interacting cloud containing the same number of atoms.  
 }
\label{FigS2}
\end{figure}

As a second method for thermometry, we exploit the density equation of state (EoS) of the unitary Fermi gas to perform a fit of the one-dimensional integrated density profile \cite{guajardo2013}. %
The density of a homogeneous unitary gas can be expressed as a universal function of $q = \beta\mu$:
\begin{equation}
n\lambda_{dB}^3 = f_n (q) = -\mbox{Li}_{3/2} (-e^q) \,\, F(q),
\end{equation}
where $\lambda_{dB} = \sqrt{2 \pi \hbar^2/m k_B T}$ is the de Broglie wavelength, with $m$ being the mass of a ${}^6$Li atom, and $F(q)= n(q)/n_0(q)$ is the ratio between the unitary and the non-interacting densities, which has been measured in Ref.~\cite{Ku2012sm} for $-0.9 \leq q \leq 3.9$. 
%, 
The EoS $f_n(q)$ can be extended out of such range of $q$ using the $4^\mathrm{th}$ order virial expansion \cite{liu2009, Nascimbene2010, Ku2012sm} for $ q < -0.9 $, and the phonon model of Ref.~\cite{taylor2009, hou2013} for $q > 3.9$. The full EoS can then be written as: 
\begin{equation}
f_n(q) =
  \begin{cases}
    \sum_{k = 1}^4 k\,b_k\,e^{kq} & \text{for } q< -0.9, \\[3mm]
     -\mbox{Li}_{3/2} (-e^q)\,F(q) & \text{for } -0.9 \leq q \leq 3.9, \\[3mm]
    \frac{4}{3\sqrt{\pi}} \left[\left(\frac{q}{\xi}\right)^{\!\!3/2}-\frac{\pi^4}{480} \left( \frac{3}{q}\right)^{\!5/2}\,\right] & \text{for } q> 3.9.
  \end{cases}
 \label{Eq:EoS_UFG}
 \bigskip
\end{equation}
Here, $b_k$ denotes the $k$-th virial coefficient, taken from Refs.~\cite{liu2009,Ku2012sm}. 
In a trapped sample, the local density approximation (LDA) can be used to write the chemical potential in each point of the cloud as:
\begin{equation}
    \mu (x, y, z) = \mu_0 - V_\mathrm{trap} (x, y, z)= \mu_0 - \frac{1}{2} m \left( \omega_x ^2 x^2 +\omega_y^2 y^2 + \omega_z^2 z^2 \right),
    \label{Eq:EoSforMu}
\end{equation}
where $\mu_0$ is the chemical potential at the trap center and $V_\mathrm{trap } (x, y, z)$ is the harmonic trapping potential. Under this assumption, the density profile of a trapped unitary Fermi gas is thus:
\begin{equation}
    n(x, y, z) = \frac{1}{\lambda_{dB}^3} \, f_n\!\left(q(x, y, z)\right) = \frac{1}{\lambda_{dB}^3} \, f_n \!\left(\beta \mu_0 -\frac{\beta m}{2} \left( \omega_x^2 x^2 +\omega_y^2 y^2 + \omega_z^2 z^2  \right) \right).
    \label{Eq:EoSinTrap}
\end{equation}
The one-dimensional density profile $n_{1D} (y)$ can be obtained by integrating the above expression along the $x$ and $z$ directions. It can be shown that this is given by:
\begin{equation}
    n_{1D} (y) = \frac{2 \pi }{m \omega_z \omega_x} \frac{k_B T}{\lambda_{dB}^3}\, f_s\!\left(\beta \mu_0 -\frac{\beta m}{2} \omega_y^2 y^2\right),
\label{Eq:fit1D_eos}
\end{equation}
where the function $f_s (q) = \int_{-\infty}^q f_n(s)\,ds$ is related to the EoS in Eq.~\eqref{Eq:EoS_UFG}. A similar relation holds for the the one-dimensional densities along the $x$ and $z$ directions. 
Using Eq.~\eqref{Eq:fit1D_eos} as a fitting function of the measured one-dimensional density profile, we can extract the degeneracy parameter $T/T_F$ of the trapped cloud. In Fig.~\ref{FigS2}(a), two typical radial density profiles for $T/T_F = 0.125(3)$ (red) and $T/T_F = 0.077(5)$ (blue) are shown together with the corresponding fits of Eq.~\eqref{Eq:fit1D_eos}.
We chose to perform the fit along the $y$-direction, since the wings of the density profile, which are essential for the temperature extraction, lie entirely within the field of view of our imaging system, in contrast to the axial $x$-direction.
%. 

As a further check of our thermometry at unitarity, we employ a method based on the virial theorem \cite{thomas2005}, that links the total energy $E$ of the cloud to the second moment of its density distribution $\langle y^2 \rangle$:
\begin{align}
E &= 3 m \omega_y^2 \,\langle y^2 \rangle,\\[2mm]
\langle y^2 \rangle &=  \int^{+\infty}_{-\infty} n_{1D}(y) \,y^2 \, dy.
\end{align}

\noindent From the in-situ images of the cloud, we extract the second moment of the density distribution and calculate $E$. From this, we obtain the temperature using the knowledge on the equation of state of the unitary Fermi gas: the total energy has a monothonic dependence on $T/T_F$, as plotted in Fig.~\ref{FigS2}(b). There, the total energy of the unitary gas $E$ is normalized to that one of a non-interacting cloud with the same number of particles at zero temperature, $E_0 = \frac{3}{4} N E_F$.

All the presented thermometry methods produce consistent temperature estimations. The values of $T/T_F$ reported in the main text are obtained from a weighted average of the three methods, while the uncertainty is set equal to $\pm 0.01$, that is the mean observed deviation among the different methods. At very low temperatures $T < 0.1\,T_F$, where the phenomenological fit is known to become most unreliable, we consider only the results of the other methods for the final temperature estimation.
When averaging measurements from different experimental realization (Figs.~2-~3 of main text), the error over $T/T_F$ is assumed to be equal to $\pm 0.01$ as well, since the uncertainty on the temperature extraction is always larger than the statistical uncertainty from the averaging.

\subsection{Thermometry of crossover Fermi gases away from unitarity}

%.
%
To produce samples in the BCS-BEC interaction crossover, we evaporate on top of the Feshbach resonance, and only at the end of evaporation we slowly sweep the magnetic field to the desired value in 50\,ms. 
Being the entropy constant during an adiabatic magnetic-field sweep, we assume the temperature of the final gas to remain the same when moving towards the BCS side of the resonance \cite{Haussmann2007sm}. 
To estimate the gas temperature with $1/k_Fa = -0.45$, we therefore extract the temperature of the unitary gas before the Feshbach field sweep using the same methods described in Sec.~\ref{Sec:SM_UFGthermo}.
On the other hand, on the BEC side of the resonance the degeneracy parameter is not expected to remain constant during an adiabatic process \cite{Haussmann2007sm, Carr2004}. A lower bound of $T/T_F$ for $1/k_Fa = 0.53$ is provided by $T/T_F$ of the unitary gas before the field sweep (and only a small deviation is expected in the regime of our measurement close to unitarity \cite{Haussmann2007sm}). We use this estimation for the data point in Fig.~2(c), without further corrections. We stress that the actual temperature of the cloud with $1/k_Fa = 0.53$ is expected to be slightly higher than our estimation \cite{Haussmann2007sm}, such that the observed monotonic increase of $T_0$ is expected to be a robust feature.

\subsection{Extraction of local quantities}

\begin{figure}[t]
\centering
\includegraphics[width = 160mm]{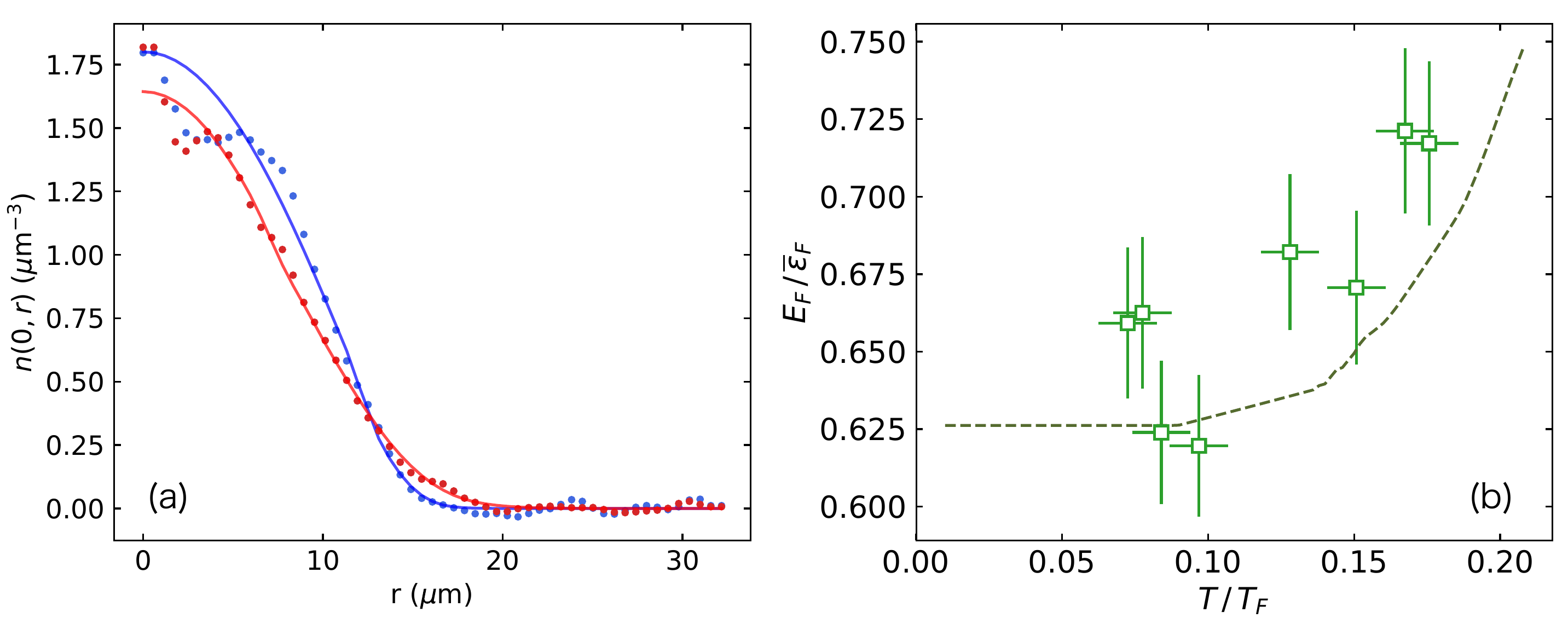}
 \caption{Extraction of local thermodynamic quantities. (a) Radial cut of the Abel-reconstructed local density profiles for unitary Fermi gases at $T/T_F = 0.08(1)$ (blue points) and $T/T_F = 0.18(1)$ (red points). Solid lines represent the density profile calculated from Eq.~\eqref{Eq:EoSinTrap} with no free parameters for the two temperatures. 
 (b) Re-scaling factor $E_F/\bar{\epsilon}_F$ from data sets at different temperatures. The re-scaling factor calculated assuming the density profile of Eq.~\eqref{Eq:EoSinTrap} is plotted as a dashed line.
 }
\label{FigS3}
\end{figure}

All thermodynamic properties introduced in the main text are based on trap-averaged quantities. It is possible to express them in terms of local quantities by accessing the local density $n(x, y, z)$. The absorption imaging technique that we use to probe the system does not give direct information on the local 3D density $n(x, y, z)$, but rather on its integral along the imaging $z$-direction, $n_{2D} (x, y)$. Under the assumption of elliptical symmetry, the 3D density profile can be reconstructed by using an inverse Abel transform of the integrated density profile:
\begin{equation}
    n (x, r) = -\frac{1}{\pi} \int_r^{+\infty} \frac{d n_{2D}(x, y)}{dy} \frac{dy}{\sqrt{y^2-r^2}}.
\end{equation}
Fig.~\ref{FigS3}(a) shows the reconstructed 3D density profile, cut along the radial direction in the center of the cloud, for $T/T_F = 0.08(1)$ (blue dots) and $T/T_F = 0.18(1)$ (red dots). We compare the reconstructed density profiles with the calculated ones using Eq.~\eqref{Eq:EoSinTrap} with no free parameters, plotted as blue and red solid line for the two temperatures, respectively.
In particular, we employ the EoS definition in Eqs.~\eqref{Eq:EoSforMu}-\eqref{Eq:EoSinTrap}
for calculating the chemical potential $\mu_0$ at finite temperature, and use the number of atoms and trap frequencies reported in Sec.~\ref{Sec:SamplePrep}.
The agreement among the experimental and calculated density profiles is particularly good on the wings of the cloud, whereas in the trap center the reconstructed profile is affected by unavoidable noise from the inverse Abel reconstruction. Such excellent agreement confirms that the density profile of our unitary cloud is well described by Eq.~\eqref{Eq:EoSinTrap}.

The local 3D density profile can be used to calculate the local Fermi energy $\epsilon_F (x, r) = \hbar^2/2 m \left( 6 \pi^2 n(x, r) \right)^{2/3}$, allowing to normalize all thermodynamic quantities presented in the main text in terms of local quantities -- most importantly to obtain the local reduced temperature $k_BT/\epsilon_F$.
In particular, a convenient re-scaling factor is given by $E_F/ \bar{\epsilon}_F$, where $\bar{\epsilon}_F$ is the density-weighted Fermi energy inside a characteristic volume $V_b$:
\begin{equation}
    \bar{\epsilon}_F = \frac{\int_{V_b} \epsilon_F (\textbf{r})\, n(\textbf{r}) \,d^3 \textbf{r}}{\int_{V_b}  n(\textbf{r}) \,d^3 \textbf{r}}.
    \label{eq:barepsilonF}
\end{equation}
In our case, a reasonable choice of $V_b$ is the volume spanned by the barrier translation. In particular, we define $V_b$ as the volume delimited by a cylinder of height $\delta x = 10 \, \mu$m  and radius $r_b = 6\, \mu$m centered around the axis of the harmonic potential; this represents a trade-off between a sufficiently small volume to express local information, and a sufficiently large volume to average out the typical noise in the Abel reconstructed profiles.
The re-scaling factors $E_F/\bar{\epsilon}_F$ obtained with such definition for datasets at different temperature are plotted in Fig.~\ref{FigS3}(b). These are compared with the calculated re-scaling curve (dashed line) from Eq.~\eqref{eq:barepsilonF} using the calculated density profiles, such as the solid lines shown in Fig.~\ref{FigS3}(a). 
%.
The agreement between experimental points and calculated profile in Fig.~\ref{FigS3} confirms that all local thermodynamic quantities in our experiment can be obtained by assuming the density profile in Eq.~\eqref{Eq:EoSinTrap}.

\heavyrulewidth=.08em
\lightrulewidth=.05em
\cmidrulewidth=.03em
\belowrulesep=.65ex
\belowbottomsep=0pt
\aboverulesep=.4ex
\abovetopsep=0pt
\cmidrulesep=\doublerulesep
\cmidrulekern=.5em
\defaultaddspace=.5em

\begin{table}
    \caption{\label{tab:fits}%
	    Extracted parameters from RCSJ model fits to the measured $I-\Delta\mu$ characteristics for three different interaction strengths $1/k_F a$ in the BCS-BEC crossover regime, used for Figs.~2 and 3 of the main text. Standard fitting errors or experimental uncertainties are given in parenthesis were applicable.%\\ %PLEASE INSERT FIT RESULTS OF ALL DATASETS. 
	    \bigskip}
    \begin{tabular}{c c c c c } 
        \midrule\midrule
        \multicolumn{5}{l}{Interaction strength $(k_Fa)^{-1} \simeq 0$, Barrier height $V_0/\mu_0 \simeq 0.7$}\\ 
        \midrule
        Temperature & Fermi energy & Charging energy & Maximum supercurrent & Conductance\\
        $T/T_F$ & $E_F$ (kHz$\times h$) & $E_c$ (Hz$\times h$) & $I_{s,\mathrm{max}}$ ($10^5$/s) & $G$ ($h^{-1}$)\\ 
        \midrule
        0.073(10) & 11.4(4) & 0.123 & 3.54(24) & 8781(710)\\
        0.078(10) & 11.2(4) & 0.126 & 4.14(31) & 9601(669)\\
        0.084(10) & 10.9(4) & 0.121 & 3.33(24) & 10268(696)\\
        0.097(10) & 10.9(4) & 0.121 & 3.33(27) & 9073(891)\\
        0.102(10) & 11.1(4) & 0.129 & 3.16(27) & 6602(1614)\\
        0.128(10) & 11.4(4) & 0.122 & 3.25(25) & 9201(852)\\
        0.148(10) & 11.5(4) & 0.12 & 1.96(12) & 5419(283)\\
        0.154(10) & 11.3(4) & 0.126 & 2.58(25) & 4118(242)\\
        0.167(10) & 11.6(4) & 0.121 & 0.12(25) & 3594(220)\\
        0.170(10) & 11.2(4) & 0.126 & 2.77(20) & 3398(217)\\
        0.175(10) & 11.3(4) & 0.125 & 0.30(14) & 3803(218)\\
        0.195(10) & 11.1(4) & 0.129 & 0.27(8) & 3532(202)\\
        0.197(10) & 10.9(4) & 0.124 & 0.48(16) & 3900(238)\\
        0.221(10) & 10.5(4) & 0.126 & 0.21(14) & 2402(210)\\
        \midrule\midrule
    \end{tabular}
    \begin{tabular}{c c c c } 
        \midrule\midrule
        \multicolumn{4}{l}{Interaction strength $(k_Fa)^{-1} \simeq -0.45$, Barrier height $V_0/\mu_0 \simeq 0.75$}\\ 
        \midrule
        Temperature & Fermi energy & Maximum supercurrent & Charging rate \\
        $T/T_F$ & $E_F$ (kHz$\times h$) & $I_{s,\mathrm{max}}$ ($10^5$/s) & $G E_c$ (Hz)\\ 
        \midrule
        0.067(10) & 9.3(3) & 2.25(50) & 407(31)\\
        0.086(10) & 10.2(4) & 1.34(23) & 231(16)\\
        0.099(10) & 9.5(3) & 1.79(30) & 421(40)\\
        0.12(10) & 9.4(3) & 1.84(79) & 310(28)\\
        0.145(10) & 10.4(4) & 1.62(11) & 238(14)\\
        0.152(10) & 10.3(4) & 1.60(9) & 347(20)\\
        0.156(10) & 9.4(3) & 0.21(28) & 535(76)\\
        0.156(10) & 10.4(4) & 0.44(40) & 206(17)\\
        0.176(10) & 9.3(3) & 0.28(42) & 225(24)\\
        0.206(10) & 10.4(4) & 0.06(27) & 376(39)\\
       \midrule \midrule
    \end{tabular}
    \begin{tabular}{c c c c } 
        \midrule\midrule
        \multicolumn{4}{l}{Interaction strength $(k_Fa)^{-1} \simeq +0.53$, Barrier height $V_0/\mu_0 \simeq 0.9$}\\ 
        \midrule
        Temperature & Fermi energy & Maximum supercurrent & Charging rate \\
        $T/T_F$ & $E_F$ (kHz$\times h$) & $I_{s,\mathrm{max}}$ ($10^5$/s) & $G E_c$ (Hz)\\ 
        \midrule
        0.112(10) & 9.7(3) & 2.28(17) & 581(48)\\
        0.148(10) & 9.7(3) & 1.70(9) & 298(18)\\
        0.163(10) & 10.6(4) & 0.34(4) & 494(53)\\
        0.170(10) & 9.5(3) & 1.21(7) & 223(14)\\
        0.187(10) & 10.4(4) & 0.21(8) & 370(32)\\
        0.195(10) & 9.5(3) & 0.87(8) & 292(21)\\
        0.198(10) & 9.3(3) & 0.96(7) & 258(19)\\
        0.206(10) & 9.5(3) & 0.43(8) & 348(26)\\
        0.207(10) & 9.0(3) & 0.29(2) & 194(13)\\
        0.228(10) & 8.9(3) & 0.18(3) & 308(24)\\
        0.232(10) & 8.5(3) & 0.007(72) & 152(15)\\ 
        \midrule \midrule
    \end{tabular}
\end{table}

\section{Experimental extraction of tunnel junction parameters}

In this section, we describe the methods used to characterize the %
response of the junction to an external current. %  
We employ the resistively and capacitively shunted junction (RCSJ) circuit model already used in Ref.~\cite{Kwon2019}. This includes three parallel elements, into which the external current $I_\mathrm{ext}$ is injected: (i) a Josephson junction with current-phase relation $I_s(\varphi)$, where $\varphi = \varphi_L - \varphi_R$ is the phase difference of the two reservoirs, (ii) a shunt resistance $R$, and (iii) a capacitance $C$.
The circuit dynamics is described by the following two coupled differential equations:
\begin{subequations}
\begin{align}
\label{Eq:circuit_Iext}
I_\mathrm{ext} &= I_{s} (\varphi) + G \Delta\mu + C\Delta \Dot{\mu},\\
\hbar \Dot{\varphi} &= - \Delta \mu,
\label{Eq:circuit_PhiDot}
\end{align}
\end{subequations}
where $G = R^{-1}$ is the bias-independent tunneling conductance. 
Equations~\eqref{Eq:circuit_Iext} and \eqref{Eq:circuit_PhiDot} can be used to fit the measured current-imbalance characteristics, by considering that the chemical potential difference $\Delta \mu = \mu_{L} -\mu_{R} $ is related to the relative number imbalance $z = (N_L-N_R)/N$ by:
\begin{equation}
    \Delta \mu = \frac{N}{2} E_c \left( z - \bar{z} \right),
    \label{Eq:circuit_Dmu}
\end{equation}
where $N$ is the total number of atoms per spin state in the system,
$E_{c} = (\partial \mu_L/\partial N_L + \partial \mu_R/\partial N_R) \simeq 2\partial \mu_L/\partial N_L$ is the charging energy of the reservoirs~\cite{Meier2001sm,Giovanazzi2000sm}, and $\bar{z}$ is the relative population imbalance at equilibrium for the final barrier position. 
By numerically solving Eqs.~\eqref{Eq:circuit_Iext}-\eqref{Eq:circuit_PhiDot}, we fit the experimental $I-\Delta \mu$ characteristics leaving only $I_{s,\mathrm{max}}$ and $G$ as free parameters (see \cite{Kwon2019} for details). In all fits, we include only the first harmonic of the current-phase relation, i.e. $I_s(\varphi) = I_{s,\mathrm{max}} \sin \varphi$ (see also Sec.~\ref{Sec:TheoIc}). Such approximation does not compromise the accuracy of fits to the measured $I-\Delta\mu$ for the typical barrier height $V_0 \gtrsim 0.7\,\mu_0$ employed here \cite{Kwon2019}.

For each set of experimental parameters (temperature, interaction strength, barrier strength), we acquire two independent data sets with $I_\mathrm{ext} >0$ and $I_\mathrm{ext} <0$, and perform a single fit, after connecting them together by constraining the response of the junction to be anti-symmetric around 0. 
The method remains valid even at high temperatures, where $I_{s,\mathrm{max}}$ vanishes, provided that an adequate finite-temperature determination of $E_c$ is used. In particular, at unitarity $E_{c}$ is evaluated through knowledge of the confining potential \cite{Eckel2016} and the density EoS in Eq.~\eqref{Eq:EoS_UFG}.
On the other hand, away from unitarity there exist no analytic expression of the EoS at finite temperatures, and the estimation of $E_c$ is not straightforward. Therefore, we only extract $G$ separately at unitarity, while we consider the tunnel conduction rate $G E_c$ directly extracted from the RCSJ fit for all measurements at $1/k_Fa \neq 0$. %  
The values of $I_{s,\mathrm{max}}$ and $G$ extracted from all fits are summarized in Table~\ref{tab:fits}.

\section{Extraction of the dc Josephson breakdown temperature}

For crossover gases at $1/k_F a = -0.45$ and $1/k_F a = 0.53$, we perform a characterization of $I_{s,\mathrm{max}}/I_F$ at varying temperature analogous to that shown in Fig.~2(b) for a unitary Fermi gas. The junction parameters obtained for all three interaction strengths are listed in Table~\ref{tab:fits}.
To extract $T_0/T_F$, namely the critical temperature for the breakdown of the dc Josephson effect, we use an empirical fit of the maximum current $\hbar I_{s,\mathrm{max}}/E_F$ obtained experimentally as a function of $T/T_F$. In particular, we use the fitting function:
\begin{equation}
    I_\text{fit}\!\left(T/T_F\right) = I_{s,0} \times \left[ 1 - \left( \frac{T/T_F}{T_0/T_F}\right)^{\!\!\!\alpha}\, \right] \times \theta \left( T/T_F - T_0/T_F\right),
\label{Eq:fit_T0ovTF}
\end{equation}
where $\theta(x)$ is the Heaviside function.  
%,
Fit results for $I_{s,0}$ and $(T_0/T_F)$ are listed in Tab.~\ref{tab:fitResults_T0ovTF}, while Fig.~2(c) in the main text displays $(T_0/T_F)$ extracted at the three interaction strengths.

\begin{table}[h]
    \begin{tabular}{c c c c } 

        \midrule\midrule
        $1/k_F a$ & $\hbar I_{s,0}/E_F$ & $T_0/T_F$\\ 
        \midrule
        $-0.45$ & 2.7(2) & 0.164(10)\\
        0 & 5.1(4) & 0.193(10)\\
        0.53 & 4.8(11) & 0.228(10)\\
        \midrule\midrule
    \end{tabular}
        \caption{\label{tab:fitResults_T0ovTF}%
	    Results of the fit with Eq.~\eqref{Eq:fit_T0ovTF} to the measured $I_{s,\mathrm{max}}$ versus $T/T_F$, for three different interaction strengths $1/k_F a$ in the BCS-BEC crossover regime. Standard fitting errors are given in parenthesis.}
\end{table}

\section{Theoretical methods}

\subsection{Theoretical model of the Josephson current}\label{Sec:TheoIc}
Here, we describe in some detail our theoretical microscopic model for the maximum Josephson supercurrent $I_{s,\mathrm{max}}$ at unitarity for finite temperatures. This is an extension of the recently developed analytic model addressing the Josephson critical current in $T=0$ crossover superfluids \cite{Zaccanti2019sm}, which has been already applied and substantiated in Ref.~\cite{Kwon2019}. In particular, this can be applied also in the presence of inhomogeneous trapping potentials via LDA~\cite{Zaccanti2019sm}. 
The Josephson supercurrent depends on the relative phase $\varphi = \varphi_L - \varphi_R$ between the pair condensates in the two reservoirs, and within a tunneling Hamiltonian description \cite{Prange1963}, it can be expressed as a Fourier series $I_s(\varphi) = \sum_1^{\infty} I_n \sin (n \varphi)$, where the coefficients $I_n$ are linked to the amplitude for $n$ pairs to coherently tunnel through the barrier \cite{Bloch1970sm}. For barrier heights $V_0 \gtrsim \mu_0$, the single-pair transmission probability $|t_p|^2$ through the junction is small, i.e., $|t_p|^2\ll 1$, and the two lowest-order contributions suffice to estimate the maximum current within a few percent \cite{Kwon2019}. 
Following Ref. \cite{Zaccanti2019sm}, % 
the first-order contribution $I_1$ to the supercurrent is given by:
\begin{equation}
\hbar I_1 = \int_{\mathcal{V}} d^3 \textbf{r} \,\lambda_{0}(\mathbf{r})\,n_{}(\mathbf{r})\,\mu(\mathbf{r})\, \frac{|t_p(\mu(\mathbf{r}),V)|}{4\,k(\mu(\mathbf{r}))\,R_x},
 \label{realIc}
\end{equation}
where $\mathcal{V}$ is the total volume of the system, $\lambda_{0}(\mathbf{r})$ is the local condensate fraction, $n_{}(\mathbf{r})$ is the local density per spin state, $R_x$ is the zero temperature axial Thomas-Fermi radius of the cloud, $k(\mu)=\sqrt{2m \mu}/\hbar$, and $|t_p(\mu,V)|$ is the transmission amplitude through a barrier of height $V$ for a single pair with mass $m_p=2m$ and energy given by the pair chemical potential $\mu=\mu_p$ (excluding the negative in-vacuum binding energy, if any). %
\begin{figure}[t]
\centering
\includegraphics[width = 90mm]{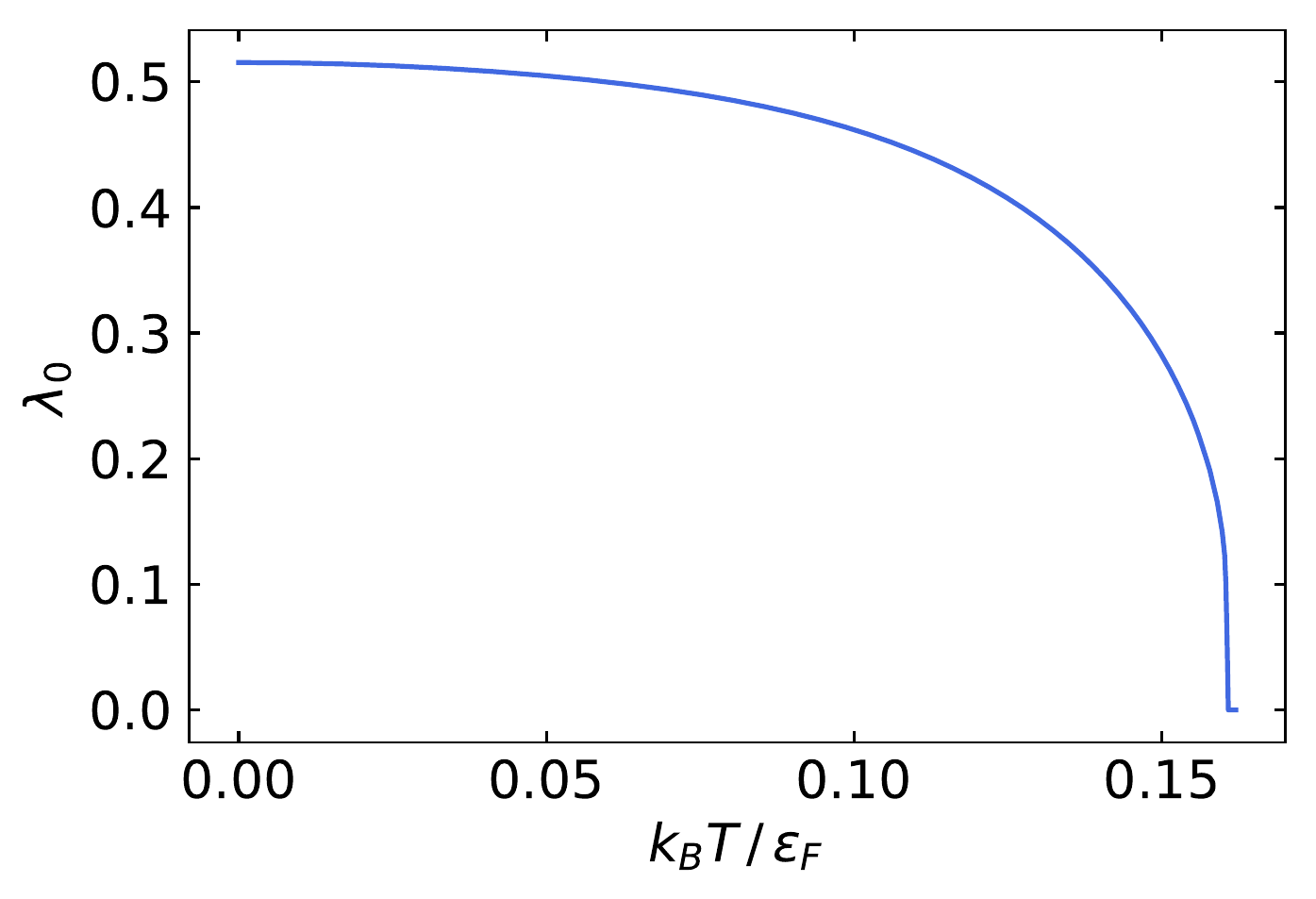}
 \caption{Condensate fraction of the homogeneous unitary Fermi gas as a function of temperature computed by the Luttinger-Ward technique \cite{Haussmann2007sm}.}
\label{FigS4}
\end{figure}
At unitarity, Eq.~\eqref{realIc} can be evaluated at finite temperatures by employing the unitary Fermi gas EoS to calculate the temperature-dependent local chemical potential $\mu(\mathbf{r})$ and density $n(\mathbf{r})$, given respectively by Eq.~\eqref{Eq:EoSforMu} and Eq.~\eqref{Eq:EoSinTrap}. Moreover, we use the condensate fraction $\lambda_{0}$ of a homogeneous unitary Fermi gas at finite temperature, obtained with a non-perturbative Luttinger-Ward technique \cite{Haussmann2007sm} (see Fig.~\ref{FigS4}). We evaluate the local condensate fraction $\lambda_0 (\textbf{r})$ in the trapped gas by taking into account the local Fermi energy $\epsilon_F(\textbf{r})$, and defining in each point of the trap the local degeneracy parameter $k_B T/ \epsilon_F (\textbf{r})$, thereby obtaining $\lambda_0 (\textbf{r}) \equiv \lambda_0 (k_B T/ \epsilon_F (\textbf{r}))$. 

To model the single-pair junction transmission, we employ the analytical expression for the tunneling amplitude of a particle through an Eckart potential barrier. The Eckart profile $V(x) = V/ \cosh ^2 (x/d)$ provides an excellent approximation of the real Gaussian barrier potential with $1/e^2$ width $w$ and height $V$, provided that the size $d$ of the Eckart barrier is set to $d \simeq 0.6\,w$, as already verified in Refs.~\cite{Zaccanti2019sm, Kwon2019}. 
The transmission probability $T = |t|^2$ for a particle of mass $M$ and energy $\epsilon$ through an Eckart potential barrier of height $V$ is given by \cite{haar}:
\begin{equation}
    T (\epsilon, V) = 
    \begin{cases}
    \frac{\sinh^2 \left( \pi \sqrt{{2\epsilon/\epsilon_b}} \right)}{\sinh^2 \left( \pi \sqrt{{2\epsilon/\epsilon_b}} \right) + \cosh^2 \left( \frac{\pi}{2} \sqrt{8 {V/\epsilon_b}-1} \right)} & \text{for } {V/\epsilon_b} \geq 1/8 \\[5mm]
    \frac{\sinh^2 \left( \pi \sqrt{{2\epsilon/\epsilon_b}} \right)}{\sinh^2 \left( \pi \sqrt{{2\epsilon/\epsilon_b}} \right) + \cos^2 \left( \frac{\pi}{2} \sqrt{1-8 {V/\epsilon_b}} \right)} & \text{for } {V/\epsilon_b} < 1/8
    \end{cases}
    \label{Eq:EckartTunnel}
    \smallskip
\end{equation}
where $\epsilon_b = \hbar^2 / M d^2$ is the characteristic energy of the barrier. Since the depth of focus of our projected Gaussian barrier with $1/e^2$ width $w_0 = 0.95 \, \mu$m is comparable to the typical radial size of the harmonically trapped cloud, the barrier divergence along the propagation direction ($z$-axis) cannot be entirely neglected. 
We thus approximate the projected barrier as a single-mode Gaussian profile, focused at the center of the cloud ($z=0$) where its waist and height are set to $w_0$ and $V_0$, respectively. Within this approximation, away from $z=0$ the barrier width and height are given by $w(z) \simeq w_0 \sqrt{1+(z/z_R)^2}$ and $V(z) \simeq V_0/\sqrt{1+ (z/z_R)^2}$, where $z_R=\pi w_{0}^2/\lambda$ is the Rayleigh length associated to $w_{0}$ and $\lambda = 532$\,nm.
Hence, we evaluate the local single-pair tunneling amplitude $|t_p(\mu(\mathbf{r}),V)|$ in Eq.~\eqref{realIc} through Eq.~\eqref{Eq:EckartTunnel} by setting $M=2m$, $\epsilon = 2\mu(\mathbf{r})$, $V = 2V(z)$ and $d = 0.6\,w(z)$.

\smallskip
To estimate the maximum critical current $I_{s,\mathrm{max}}$ we also consider the second-harmonic contribution $I_2 \sin{2\varphi}$, that increases $I_{s,\mathrm{max}}$ above the value of the first-order contribution $I_1$ in Eq.~\eqref{realIc}. As shown in Ref.~\cite{Meier2001sm, Zaccanti2019sm}, the magnitude of $I_2$ can be evaluated by substituting $|t|^{2}/16$ for $|t|/4$ in Eq.~\eqref{realIc} to give: 
\begin{equation}
\hbar |I_2| = \int_{\mathcal{V}} d^3 \textbf{r} \,\lambda_{0}(\mathbf{r})\,n_{}(\mathbf{r})\,\mu(\mathbf{r})\, \frac{|t(\mu(\mathbf{r}),V)|^{2}}{16\,k(\mu(\mathbf{r}))\,R_x} \,
 \label{2nd}.
\end{equation} 
The maximum Josephson critical current $I_{s,\mathrm{max}}$ can then be analytically computed by following Ref.~\citenum{goldobin}:
\begin{equation}
I_{s,\mathrm{max}} = \frac{(\sqrt{1+32g^2}+3)^{3/2} (\sqrt{1+32g^2}-1)^{1/2}}{32|g|}\,I_1,
 \label{Imax}
\end{equation}
where $g\equiv |I_2|/I_1$. From Eq.~\eqref{Imax}, we evaluate $I_{s,\mathrm{max}}$ as a function of temperature, which is plotted as the shaded red curve in Fig.~2(b) of the main text.

\subsection{Theoretical evaluation of the ideal Fermi gas tunneling conductance}
\label{GidealFermi}

The normal-state bias-independent conductance $G_n$ of a tunnel junction simply corresponds to the junction conductance in the case of ideal fermionic reservoirs, and has been experimentally estimated by probing the dc response of an ideal non-interacting Fermi gas. In order to theoretically evaluate $G_n$ within our trapped, spatially inhomogeneous setup, we proceed as follows. 
For a three-dimensional homogeneous Fermi gas of spinless neutral particles, the current density flowing through a tunnel junction, across which a chemical potential imbalance $\Delta \mu$ is established, is evaluated at zero temperature as \cite{Duke1969b}:
\begin{eqnarray}
j_{n} =\frac{1}{h} \, \frac{m}{2 \pi \hbar^2} \, \int_{E_F-\Delta \mu}^{E_F} d E \, \int_{0}^{E} dE_{||} |t(E-E_{||}, V_0)|^2 
\label{jN}
\end{eqnarray}
Here $E_F$ is the bulk Fermi energy at $\Delta \mu=0$, $|t(\epsilon; V_0)|^2$ is the transmission coefficient of a single particle tunneling across a barrier of height $V_0$ (and generic shape) at energy $\epsilon$, and $m/2 \pi \hbar^2$ is the two-dimensional density of states per unit area. 

In the limit of $\Delta \mu \rightarrow 0$, i.e., within linear response, Eq.~\eqref{jN} yields a current density which scales linearly with the chemical potential imbalance,
\begin{eqnarray}
j_{n} =\frac{1}{h} \, \frac{m}{2 \pi \hbar^2} \, \int_{0}^{E_F} dE_{||} |t(E_F-E_{||}, V_0)|^2 \, \Delta \mu ,
\label{jN2}
\end{eqnarray}
from which the normal-state bias-independent conductance per unit area is readily evaluated as:
\begin{eqnarray}
hg_{n} = \frac{h j_n}{\Delta \mu} =\frac{m}{2 \pi \hbar^2} \, \int_{0}^{E_F} dE_{||} |t(E_F-E_{||}, V_0)|^2.
\label{gN}
\end{eqnarray}
Physically, the integral in Eq.~\eqref{gN} over $E_{||} \in [0, E_F]$ accounts for all available transverse modes within a three-dimensional Fermi gas, bisected by an extended tunnel junction \cite{Duke1969b}.

The total conductance in our trapped configuration is evaluated on the basis of Eq. (\ref{gN}) within the local density approximation. At each position $(r,z=0)$, where the gas at density $n_F(r,0)$ is characterized by a local Fermi energy $\epsilon_F(r,0)= \frac{\hbar^2}{2 m} (6 \pi^2 n_F(r,0))^{2/3}$, % 
the conductance per unit area is given by:
\begin{eqnarray}
hg_{n} (r)= \frac{m}{2 \pi \hbar^2} \, \int_{0}^{\epsilon_F(r,0)} dE_{||} |t(\epsilon_F(r,0)-E_{||}, V)|^2.
\label{gNr}
\end{eqnarray}
The total conductance of our junction is then obtained upon integrating over the transverse directions as
\begin{eqnarray}
G_{n}= \int_{0}^{R_{F, \bot}} 2 \pi r \, g_n (r)\,dr,
\label{GN}
\end{eqnarray}
with $R_{F, \bot}$ denoting the transverse (radial) Thomas-Fermi radius of the trapped sample.
To evaluate $G_n$, we employ the Eckart barrier approximation to calculate the transmission coefficient $|t(\epsilon, V)|^2$ and account for the Gaussian divergence of the barrier profile along its propagation direction, as done for the maximum supercurrent calculation. %, 
Finally, we remark that the conductance given by Eq.~(\ref{GN}), that strictly holds only at $T=0$, is marginally affected by finite temperature effects, as long as $k_B T \ll E_F, V_0$. Indeed, the calculated $G_n = 160(29)\,h^{-1}$ is found in reasonable agreement with the conductance $G_0 = 102(15)\,h^{-1}$ measured for an ideal (spin-polarized) Fermi gas at $T/T_F \simeq 0.21$ tunneling through a barrier of height $V_0 \simeq 0.73\,E_F$. The uncertainty on the calculated $G_n$ results from a tolerance of $\pm5\%$ over the nominal Eckart barrier width $d$.

\subsection{Scaling of the tunneling conductance with adimensional barrier strength}

In a perturbative bosonic description of the tunnel junction, neglecting therefore pair breaking processes, the normal current $I_n = G \Delta \mu$ arises only at second-order in the tunnel coupling between the pair condensates \cite{Meier2001sm,Uchino2019sm}. The conductance $G$ scales thus as $G \sim |t_p|^2$, where $|t_p|$ is the single-pair tunneling amplitude. 
Such relationship is expected to break down in our experiment once a broken-pair normal phase emerges above the superfluid critical temperature. There, an equivalent scaling of the normal-state conductance $G_n \sim |t_F|^2$ holds in terms of the tunneling amplitude $|t_F|$ of a single fermion: in the absence of sizable dissipation within the reservoirs and in the small-bias linear response regime, the tunneling conductance reflects essentially the fermion transmission probability $|t_F|^2$ \cite{Duke1969b} (see Sec.~\ref{GidealFermi}).

\begin{figure}[b]
\centering
\includegraphics[width=90mm]{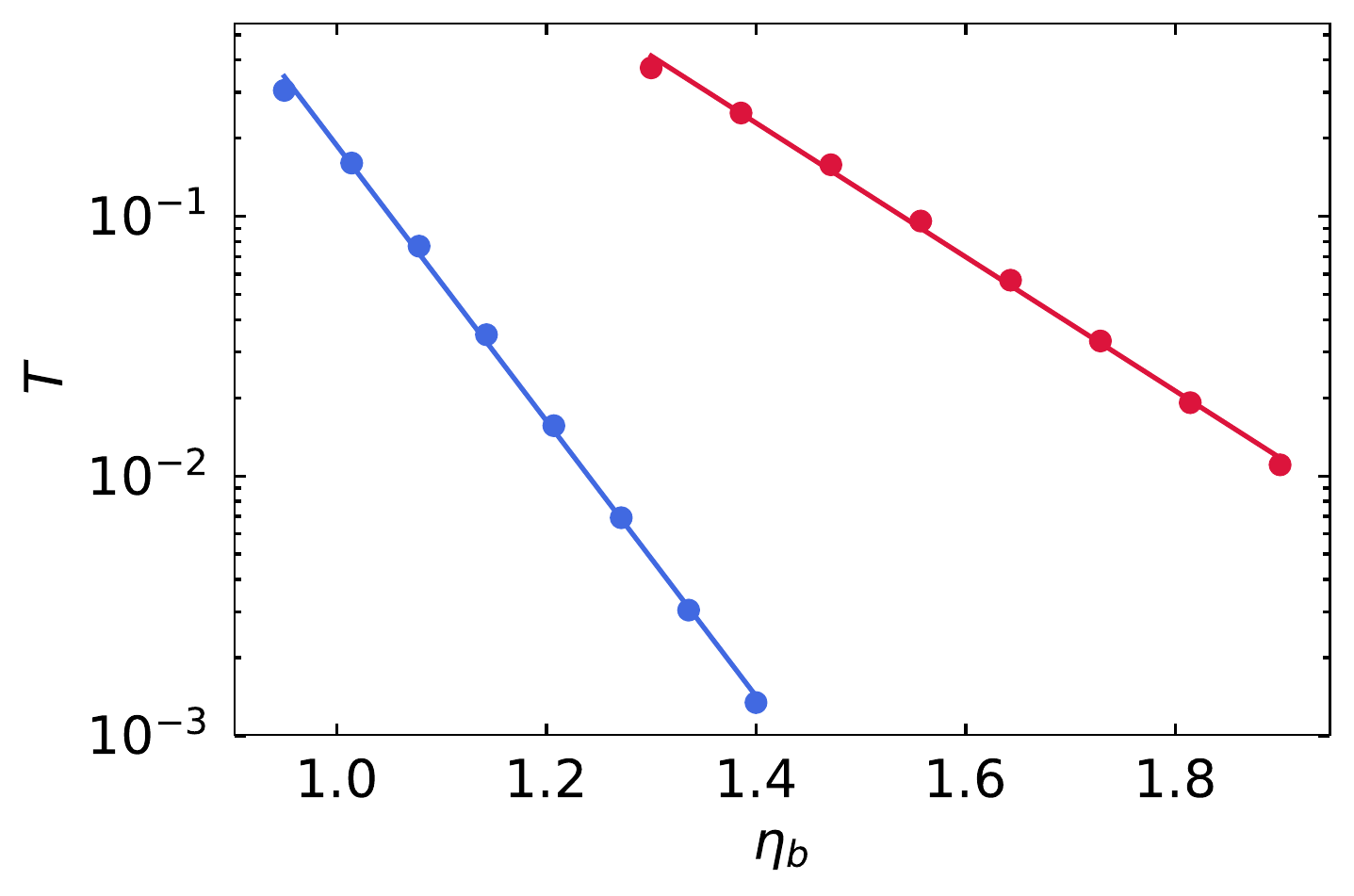}
 \caption{Tunneling probability through an Eckart barrier with dimension $d = 0.6 \times 0.95 \, \mu$m as a function of the adimensional barrier strength $\eta_b$. $T$ is calculated following Eq.~\eqref{Eq:EckartTunnel} for a particle of energy $\epsilon = h \sqrt{\xi} \times 6$\,kHz and mass $2 m$ (blue circles), and for a particle of energy $\epsilon = h \sqrt{\xi} \times 11$\,kHz and mass $m$ (red circles). In both cases, the experimental conditions and the plotted range of $\eta_b$ are the same as in Fig.~4. Solid lines represent linear fits of $\log T$. The ratio of the angular coefficient of the blue and red line is $2.06 \pm 0.07$.
 }\vspace*{-10pt}
\label{FigS5}
\end{figure}

As already discussed, an Eckart potential barrier well approximates the real junction potential, and the single-particle transmission probability $T(\epsilon) = |t(\epsilon)|^2$ (for a pair or a fermion at energy $\epsilon$) can be computed analytically by Eq.~\eqref{Eq:EckartTunnel}.  In the tunneling regime where $V/\epsilon_b \gg 1/8$ and $\epsilon/\epsilon_b > 1$, Eq.~\eqref{Eq:EckartTunnel} can be expanded as:
\begin{equation}
    |t(\epsilon)|^2 \sim \exp{ \left(2 \pi \sqrt{ \frac{2\epsilon}{\epsilon_b}}\right)} \times \exp{\left(- 2 \pi \sqrt{ \frac{2V}{\epsilon_b}} \right)}.
\end{equation}
Taking the logarithm of the tunneling probability, we obtain:
\begin{equation}
    \log |t(\epsilon)|^2 \sim - 2 \pi \sqrt{\frac{M}{m}} (k_F d) \left( \sqrt{\frac{V}{E_F}} - \sqrt{\frac{\epsilon}{E_F}}\right),
    \label{eq:t2expansion}
\end{equation}
once energies are re-scaled in units of the Fermi energy $E_F = \hbar^2 k_F^2/2 m$, where $m$ is the mass of a $^6$Li atom.
The value of $\log |t|^2$ is expected to change by a factor of about $2$ whether the tunneling particles are single fermionic atoms or pairs. Indeed, in the case of a tightly bound pair, Eq.~\eqref{eq:t2expansion} must be evaluated setting $M = m_p = 2 m$ and $\epsilon = 2 \mu$, whereas $M = m$ and $\epsilon = \mu$ apply to the single-atom case. Moreover, the potential height felt by a pair is approximately twice that felt by a single fermion, namely $V_{p} \simeq 2 V$, since the polarizability of a tightly bound pair is twice that of a single atom. Therefore, $\log |t_p|^2 \simeq 2 \log |t_F|^2$ is expected to hold in general as long as pairs are sufficiently bound objects. A similar factor of approximately $2$ is expected to appear in the scaling of the conductance, whether the conducting particles are single fermionic atoms or pairs, since $G \sim |t|^2$.

In the experimental data analysis, to investigate the scaling of the conductance with the barrier parameters, we consider the adimensional barrier strength $\eta_b = (k_F d) \times \sqrt{V_0/E_F}$. The scaling is experimentally obtained by varying the barrier height $V_0$ while keeping the barrier size $d$ constant. We expect $G$ to scale exponentially with $\eta_b$ since $|t|^2 \sim e^{-\eta_b}$ [see Eq.~\eqref{eq:t2expansion}].
In Fig.~\ref{FigS5}, we report the tunneling probability computed numerically for an Eckart barrier under the same experimental conditions of Fig.~4(a) for blue circles and Fig.~4(b) for red circles. In the first case, we assume the tunnel conduction to be mediated by pairs of mass $2m$, while in the second one by single atoms of mass $m$. In the same range of values of barrier strength $\eta_b$ of Fig.~4, the scaling of $\log |t|^2$ is observed to change drastically in the two cases. To be more quantitative, we perform a linear fit of $\log |t|^2$ and extract the slope for both cases. A ratio of $2.06 \pm 0.07$ among the pair and single-atom $\log |t|^2$ slopes is found, consistent with the factor of $2$ expected from Eq.~\eqref{eq:t2expansion}. Such calculation confirms that, under the experimental conditions of Fig.~4 of the main text, the scaling of $\log G$ is expected to differ by a factor $\approx 2$ whether bound pairs or single atoms are the mediators of incoherent tunneling currents, and elucidates the observed scaling of $\log G$ versus $\eta_b$ well below, around and above $T_c$.

\vspace*{-10pt}


\begin{thebibliography}{52}%
\makeatletter
\providecommand \@ifxundefined [1]{%
 \@ifx{#1\undefined}
}%
\providecommand \@ifnum [1]{%
 \ifnum #1\expandafter \@firstoftwo
 \else \expandafter \@secondoftwo
 \fi
}%
\providecommand \@ifx [1]{%
 \ifx #1\expandafter \@firstoftwo
 \else \expandafter \@secondoftwo
 \fi
}%
\providecommand \natexlab [1]{#1}%
\providecommand \enquote  [1]{``#1''}%
\providecommand \bibnamefont  [1]{#1}%
\providecommand \bibfnamefont [1]{#1}%
\providecommand \citenamefont [1]{#1}%
\providecommand \href@noop [0]{\@secondoftwo}%
\providecommand \href [0]{\begingroup \@sanitize@url \@href}%
\providecommand \@href[1]{\@@startlink{#1}\@@href}%
\providecommand \@@href[1]{\endgroup#1\@@endlink}%
\providecommand \@sanitize@url [0]{\catcode `\\12\catcode `\$12\catcode
  `\&12\catcode `\#12\catcode `\^12\catcode `\_12\catcode `\%12\relax}%
\providecommand \@@startlink[1]{}%
\providecommand \@@endlink[0]{}%
\providecommand \url  [0]{\begingroup\@sanitize@url \@url }%
\providecommand \@url [1]{\endgroup\@href {#1}{\urlprefix }}%
\providecommand \urlprefix  [0]{URL }%
\providecommand \Eprint [0]{\href }%
\providecommand \doibase [0]{http://dx.doi.org/}%
\providecommand \selectlanguage [0]{\@gobble}%
\providecommand \bibinfo  [0]{\@secondoftwo}%
\providecommand \bibfield  [0]{\@secondoftwo}%
\providecommand \translation [1]{[#1]}%
\providecommand \BibitemOpen [0]{}%
\providecommand \bibitemStop [0]{}%
\providecommand \bibitemNoStop [0]{.\EOS\space}%
\providecommand \EOS [0]{\spacefactor3000\relax}%
\providecommand \BibitemShut  [1]{\csname bibitem#1\endcsname}%
\let\auto@bib@innerbib\@empty
%</preamble>
\bibitem [{\citenamefont {Burstein}\ and\ \citenamefont
  {Lundqvist}(1969)}]{Burstein1969}%
  \BibitemOpen
  \bibfield  {author} {\bibinfo {author} {\bibfnamefont {E.}~\bibnamefont
  {Burstein}}\ and\ \bibinfo {author} {\bibfnamefont {S.}~\bibnamefont
  {Lundqvist}},\ }\href@noop {} {\emph {\bibinfo {title} {Tunneling Phenomena
  in Solids}}}\ (\bibinfo  {publisher} {Springer, Boston MA},\ \bibinfo {year}
  {1969})\BibitemShut {NoStop}%
\bibitem [{\citenamefont {Giaever}(1960)}]{Giaever1960}%
  \BibitemOpen
  \bibfield  {author} {\bibinfo {author} {\bibfnamefont {I.}~\bibnamefont
  {Giaever}},\ }\bibfield  {title} {\bibinfo {title} {\emph {Energy Gap in
  Superconductors Measured by Electron Tunneling}},\ }\href {\doibase
  10.1103/PhysRevLett.5.147} {\bibfield  {journal} {\bibinfo  {journal} {Phys.
  Rev. Lett.}\ }\textbf {\bibinfo {volume} {5}},\ \bibinfo {pages} {147}
  (\bibinfo {year} {1960})}\BibitemShut {NoStop}%
\bibitem [{\citenamefont {Datta}(1997)}]{DattaBook}%
  \BibitemOpen
  \bibfield  {author} {\bibinfo {author} {\bibfnamefont {S.}~\bibnamefont
  {Datta}},\ }\href@noop {} {\emph {\bibinfo {title} {Electronic transport in
  mesoscopic systems}}}\ (\bibinfo  {publisher} {Cambridge University Press},\
  \bibinfo {year} {1997})\BibitemShut {NoStop}%
\bibitem [{\citenamefont {Pines}\ and\ \citenamefont
  {Nozi\'eres}(1999)}]{PinesNozieres}%
  \BibitemOpen
  \bibfield  {author} {\bibinfo {author} {\bibfnamefont {D.}~\bibnamefont
  {Pines}}\ and\ \bibinfo {author} {\bibfnamefont {P.}~\bibnamefont
  {Nozi\'eres}},\ }\href@noop {} {\emph {\bibinfo {title} {Theory of quantum
  liquids}}},\ Vol.~\bibinfo {volume} {1}\ (\bibinfo  {publisher} {Perseus
  Books Publishing, Cambridge, MA},\ \bibinfo {year} {1999})\BibitemShut
  {NoStop}%
\bibitem [{\citenamefont {Bogoliubov}\ \emph {et~al.}(1958)\citenamefont
  {Bogoliubov}, \citenamefont {Tolmachov},\ and\ \citenamefont {{\v
  S}irkov}}]{Bogoliubov1958}%
  \BibitemOpen
  \bibfield  {author} {\bibinfo {author} {\bibfnamefont {N.~N.}\ \bibnamefont
  {Bogoliubov}}, \bibinfo {author} {\bibfnamefont {V.~V.}\ \bibnamefont
  {Tolmachov}}, \ and\ \bibinfo {author} {\bibfnamefont {D.~V.}\ \bibnamefont
  {{\v S}irkov}},\ }\bibfield  {title} {\bibinfo {title} {\emph {A New Method
  in the Theory of Superconductivity}},\ }\href {\doibase
  10.1002/prop.19580061102} {\bibfield  {journal} {\bibinfo  {journal}
  {Fortschritte der Phys.}\ }\textbf {\bibinfo {volume} {6}},\ \bibinfo {pages}
  {605} (\bibinfo {year} {1958})}\BibitemShut {NoStop}%
\bibitem [{\citenamefont {Kivelson}\ and\ \citenamefont
  {Rokhsar}(1990)}]{Kivelson1990}%
  \BibitemOpen
  \bibfield  {author} {\bibinfo {author} {\bibfnamefont {S.~A.}\ \bibnamefont
  {Kivelson}}\ and\ \bibinfo {author} {\bibfnamefont {D.~S.}\ \bibnamefont
  {Rokhsar}},\ }\bibfield  {title} {\bibinfo {title} {\emph {Bogoliubov
  quasiparticles, spinons, and spin-charge decoupling in superconductors}},\
  }\href {\doibase 10.1103/PhysRevB.41.11693} {\bibfield  {journal} {\bibinfo
  {journal} {Phys. Rev. B}\ }\textbf {\bibinfo {volume} {41}},\ \bibinfo
  {pages} {11693} (\bibinfo {year} {1990})}\BibitemShut {NoStop}%
\bibitem [{\citenamefont {Anderson}(1958)}]{Anderson1958}%
  \BibitemOpen
  \bibfield  {author} {\bibinfo {author} {\bibfnamefont {P.~W.}\ \bibnamefont
  {Anderson}},\ }\bibfield  {title} {\bibinfo {title} {\emph {Random-Phase
  Approximation in the Theory of Superconductivity}},\ }\href {\doibase
  10.1103/PhysRev.112.1900} {\bibfield  {journal} {\bibinfo  {journal} {Phys.
  Rev.}\ }\textbf {\bibinfo {volume} {112}},\ \bibinfo {pages} {1900} (\bibinfo
  {year} {1958})}\BibitemShut {NoStop}%
\bibitem [{\citenamefont {Combescot}\ \emph {et~al.}(2006)\citenamefont
  {Combescot}, \citenamefont {Kagan},\ and\ \citenamefont
  {Stringari}}]{Combescot2006}%
  \BibitemOpen
  \bibfield  {author} {\bibinfo {author} {\bibfnamefont {R.}~\bibnamefont
  {Combescot}}, \bibinfo {author} {\bibfnamefont {M.~Y.}\ \bibnamefont
  {Kagan}}, \ and\ \bibinfo {author} {\bibfnamefont {S.}~\bibnamefont
  {Stringari}},\ }\bibfield  {title} {\bibinfo {title} {\emph {Collective mode
  of homogeneous superfluid Fermi gases in the BEC-BCS crossover}},\ }\href
  {\doibase 10.1103/PhysRevA.74.042717} {\bibfield  {journal} {\bibinfo
  {journal} {Phys. Rev. A}\ }\textbf {\bibinfo {volume} {74}},\ \bibinfo
  {pages} {042717} (\bibinfo {year} {2006})}\BibitemShut {NoStop}%
\bibitem [{\citenamefont {Zwerger}(2012)}]{ZwergerBook}%
  \BibitemOpen
  \bibinfo {editor} {\bibfnamefont {W.}~\bibnamefont {Zwerger}},\ ed.,\ \href
  {\doibase 10.1007/978-3-642-21978-8} {\emph {\bibinfo {title} {The BCS-BEC
  Crossover and the Unitary Fermi Gas}}},\ Vol.\ \bibinfo {volume} {836}\
  (\bibinfo  {publisher} {Springer, Berlin-Heidelberg},\ \bibinfo {year}
  {2012})\BibitemShut {NoStop}%
\bibitem [{\citenamefont {Goldstone}(1961)}]{Goldstone1961}%
  \BibitemOpen
  \bibfield  {author} {\bibinfo {author} {\bibfnamefont {J.}~\bibnamefont
  {Goldstone}},\ }\bibfield  {title} {\bibinfo {title} {\emph {Field theories
  with superconductor solutions}},\ }\href {\doibase 10.1007/BF02812722}
  {\bibfield  {journal} {\bibinfo  {journal} {Nuovo Cimento}\ }\textbf
  {\bibinfo {volume} {19}},\ \bibinfo {pages} {154} (\bibinfo {year}
  {1961})}\BibitemShut {NoStop}%
\bibitem [{\citenamefont {Landau}(1941)}]{Landau1941}%
  \BibitemOpen
  \bibfield  {author} {\bibinfo {author} {\bibfnamefont {L.}~\bibnamefont
  {Landau}},\ }\bibfield  {title} {\bibinfo {title} {\emph {Theory of the
  Superfluidity of Helium II}},\ }\href {\doibase 10.1103/PhysRev.60.356}
  {\bibfield  {journal} {\bibinfo  {journal} {Phys. Rev.}\ }\textbf {\bibinfo
  {volume} {60}},\ \bibinfo {pages} {356} (\bibinfo {year} {1941})}\BibitemShut
  {NoStop}%
\bibitem [{\citenamefont {Blonder}\ \emph {et~al.}(1982)\citenamefont
  {Blonder}, \citenamefont {Tinkham},\ and\ \citenamefont
  {Klapwijk}}]{Blonder1982}%
  \BibitemOpen
  \bibfield  {author} {\bibinfo {author} {\bibfnamefont {G.~E.}\ \bibnamefont
  {Blonder}}, \bibinfo {author} {\bibfnamefont {M.}~\bibnamefont {Tinkham}}, \
  and\ \bibinfo {author} {\bibfnamefont {T.~M.}\ \bibnamefont {Klapwijk}},\
  }\bibfield  {title} {\bibinfo {title} {\emph {Transition from metallic to
  tunneling regimes in superconducting microconstrictions: Excess current,
  charge imbalance, and supercurrent conversion}},\ }\href {\doibase
  10.1103/PhysRevB.25.4515} {\bibfield  {journal} {\bibinfo  {journal} {Phys.
  Rev. B}\ }\textbf {\bibinfo {volume} {25}},\ \bibinfo {pages} {4515}
  (\bibinfo {year} {1982})}\BibitemShut {NoStop}%
\bibitem [{\citenamefont {Tinkham}(1996)}]{Tinkham}%
  \BibitemOpen
  \bibfield  {author} {\bibinfo {author} {\bibfnamefont {M.}~\bibnamefont
  {Tinkham}},\ }\href@noop {} {\emph {\bibinfo {title} {Introduction to
  Superconductivity}}},\ \bibinfo {edition} {2nd}\ ed.\ (\bibinfo  {publisher}
  {McGraw-Hill},\ \bibinfo {address} {New York},\ \bibinfo {year}
  {1996})\BibitemShut {NoStop}%
\bibitem [{\citenamefont {Meier}\ and\ \citenamefont
  {Zwerger}(2001)}]{Meier2001}%
  \BibitemOpen
  \bibfield  {author} {\bibinfo {author} {\bibfnamefont {F.}~\bibnamefont
  {Meier}}\ and\ \bibinfo {author} {\bibfnamefont {W.}~\bibnamefont
  {Zwerger}},\ }\bibfield  {title} {\bibinfo {title} {\emph {Josephson
  tunneling between weakly interacting Bose-Einstein condensates}},\ }\href
  {\doibase 10.1103/PhysRevA.64.033610} {\bibfield  {journal} {\bibinfo
  {journal} {Phys. Rev. A}\ }\textbf {\bibinfo {volume} {64}},\ \bibinfo
  {pages} {033610} (\bibinfo {year} {2001})}\BibitemShut {NoStop}%
\bibitem [{\citenamefont {Uchino}(2020)}]{Uchino2020}%
  \BibitemOpen
  \bibfield  {author} {\bibinfo {author} {\bibfnamefont {S.}~\bibnamefont
  {Uchino}},\ }\bibfield  {title} {\bibinfo {title} {\emph {Role of
  Nambu-Goldstone modes in the fermionic-superfluid point contact}},\ }\href
  {\doibase 10.1103/PhysRevResearch.2.023340} {\bibfield  {journal} {\bibinfo
  {journal} {Phys. Rev. Res.}\ }\textbf {\bibinfo {volume} {2}},\ \bibinfo
  {pages} {023340} (\bibinfo {year} {2020})}\BibitemShut {NoStop}%
\bibitem [{\citenamefont {Krinner}\ \emph {et~al.}(2017)\citenamefont
  {Krinner}, \citenamefont {Esslinger},\ and\ \citenamefont
  {Brantut}}]{Krinner2017}%
  \BibitemOpen
  \bibfield  {author} {\bibinfo {author} {\bibfnamefont {S.}~\bibnamefont
  {Krinner}}, \bibinfo {author} {\bibfnamefont {T.}~\bibnamefont {Esslinger}},
  \ and\ \bibinfo {author} {\bibfnamefont {J.-P.}\ \bibnamefont {Brantut}},\
  }\bibfield  {title} {\bibinfo {title} {\emph {Two-terminal transport
  measurements with cold atoms}},\ }\href {\doibase 10.1088/1361-648x/aa74a1}
  {\bibfield  {journal} {\bibinfo  {journal} {J. Phys. Cond. Mat.}\ }\textbf
  {\bibinfo {volume} {29}},\ \bibinfo {pages} {343003} (\bibinfo {year}
  {2017})}\BibitemShut {NoStop}%
\bibitem [{\citenamefont {Hoinka}\ \emph {et~al.}(2017)\citenamefont {Hoinka},
  \citenamefont {Dyke}, \citenamefont {Lingham}, \citenamefont {Kinnunen},
  \citenamefont {Bruun},\ and\ \citenamefont {Vale}}]{Hoinka2017}%
  \BibitemOpen
  \bibfield  {author} {\bibinfo {author} {\bibfnamefont {S.}~\bibnamefont
  {Hoinka}}, \bibinfo {author} {\bibfnamefont {P.}~\bibnamefont {Dyke}},
  \bibinfo {author} {\bibfnamefont {M.~G.}\ \bibnamefont {Lingham}}, \bibinfo
  {author} {\bibfnamefont {J.~J.}\ \bibnamefont {Kinnunen}}, \bibinfo {author}
  {\bibfnamefont {G.~M.}\ \bibnamefont {Bruun}}, \ and\ \bibinfo {author}
  {\bibfnamefont {C.~J.}\ \bibnamefont {Vale}},\ }\bibfield  {title} {\bibinfo
  {title} {\emph {Goldstone mode and pair-breaking excitations in atomic Fermi
  superfluids}},\ }\href {https://doi.org/10.1038/nphys4187} {\bibfield
  {journal} {\bibinfo  {journal} {Nature Phys.}\ }\textbf {\bibinfo {volume}
  {13}},\ \bibinfo {pages} {943} (\bibinfo {year} {2017})}\BibitemShut
  {NoStop}%
\bibitem [{\citenamefont {Kuhn}\ \emph {et~al.}(2020)\citenamefont {Kuhn},
  \citenamefont {Hoinka}, \citenamefont {Herrera}, \citenamefont {Dyke},
  \citenamefont {Kinnunen}, \citenamefont {Bruun},\ and\ \citenamefont
  {Vale}}]{Kuhn2020}%
  \BibitemOpen
  \bibfield  {author} {\bibinfo {author} {\bibfnamefont {C.~C.~N.}\
  \bibnamefont {Kuhn}}, \bibinfo {author} {\bibfnamefont {S.}~\bibnamefont
  {Hoinka}}, \bibinfo {author} {\bibfnamefont {I.}~\bibnamefont {Herrera}},
  \bibinfo {author} {\bibfnamefont {P.}~\bibnamefont {Dyke}}, \bibinfo {author}
  {\bibfnamefont {J.~J.}\ \bibnamefont {Kinnunen}}, \bibinfo {author}
  {\bibfnamefont {G.~M.}\ \bibnamefont {Bruun}}, \ and\ \bibinfo {author}
  {\bibfnamefont {C.~J.}\ \bibnamefont {Vale}},\ }\bibfield  {title} {\bibinfo
  {title} {\emph {High-Frequency Sound in a Unitary Fermi Gas}},\ }\href
  {\doibase 10.1103/PhysRevLett.124.150401} {\bibfield  {journal} {\bibinfo
  {journal} {Phys. Rev. Lett.}\ }\textbf {\bibinfo {volume} {124}},\ \bibinfo
  {pages} {150401} (\bibinfo {year} {2020})}\BibitemShut {NoStop}%
\bibitem [{\citenamefont {Patel}\ \emph {et~al.}()\citenamefont {Patel},
  \citenamefont {Yan}, \citenamefont {Mukherjee}, \citenamefont {Fletcher},
  \citenamefont {Struck},\ and\ \citenamefont {Zwierlein}}]{Patel2019}%
  \BibitemOpen
  \bibfield  {author} {\bibinfo {author} {\bibfnamefont {P.~B.}\ \bibnamefont
  {Patel}}, \bibinfo {author} {\bibfnamefont {Z.}~\bibnamefont {Yan}}, \bibinfo
  {author} {\bibfnamefont {B.}~\bibnamefont {Mukherjee}}, \bibinfo {author}
  {\bibfnamefont {R.~J.}\ \bibnamefont {Fletcher}}, \bibinfo {author}
  {\bibfnamefont {J.}~\bibnamefont {Struck}}, \ and\ \bibinfo {author}
  {\bibfnamefont {M.~W.}\ \bibnamefont {Zwierlein}},\ }\href@noop {} {\bibinfo
  {title} {\emph {Universal Sound Diffusion in a Strongly Interacting Fermi
  Gas}}},\ \Eprint {http://arxiv.org/abs/1909.02555} {arXiv:1909.02555}
  \BibitemShut {NoStop}%
\bibitem [{\citenamefont {Stadler}\ \emph {et~al.}(2012)\citenamefont
  {Stadler}, \citenamefont {Krinner}, \citenamefont {Meineke}, \citenamefont
  {Brantut},\ and\ \citenamefont {Esslinger}}]{Stadler2012}%
  \BibitemOpen
  \bibfield  {author} {\bibinfo {author} {\bibfnamefont {D.}~\bibnamefont
  {Stadler}}, \bibinfo {author} {\bibfnamefont {S.}~\bibnamefont {Krinner}},
  \bibinfo {author} {\bibfnamefont {J.}~\bibnamefont {Meineke}}, \bibinfo
  {author} {\bibfnamefont {J.-P.}\ \bibnamefont {Brantut}}, \ and\ \bibinfo
  {author} {\bibfnamefont {T.}~\bibnamefont {Esslinger}},\ }\bibfield  {title}
  {\bibinfo {title} {\emph {Observing the drop of resistance in the flow of a
  superfluid Fermi gas}},\ }\href {\doibase 10.1038/nature11613} {\bibfield
  {journal} {\bibinfo  {journal} {Nature}\ }\textbf {\bibinfo {volume} {491}},\
  \bibinfo {pages} {736} (\bibinfo {year} {2012})}\BibitemShut {NoStop}%
\bibitem [{\citenamefont {Valtolina}\ \emph {et~al.}(2015)\citenamefont
  {Valtolina}, \citenamefont {Burchianti}, \citenamefont {Amico}, \citenamefont
  {Neri}, \citenamefont {Xhani}, \citenamefont {Seman}, \citenamefont
  {Trombettoni}, \citenamefont {Smerzi}, \citenamefont {Zaccanti},
  \citenamefont {Inguscio},\ and\ \citenamefont {Roati}}]{Valtolina2015}%
  \BibitemOpen
  \bibfield  {author} {\bibinfo {author} {\bibfnamefont {G.}~\bibnamefont
  {Valtolina}}, \bibinfo {author} {\bibfnamefont {A.}~\bibnamefont
  {Burchianti}}, \bibinfo {author} {\bibfnamefont {A.}~\bibnamefont {Amico}},
  \bibinfo {author} {\bibfnamefont {E.}~\bibnamefont {Neri}}, \bibinfo {author}
  {\bibfnamefont {K.}~\bibnamefont {Xhani}}, \bibinfo {author} {\bibfnamefont
  {J.~A.}\ \bibnamefont {Seman}}, \bibinfo {author} {\bibfnamefont
  {A.}~\bibnamefont {Trombettoni}}, \bibinfo {author} {\bibfnamefont
  {A.}~\bibnamefont {Smerzi}}, \bibinfo {author} {\bibfnamefont
  {M.}~\bibnamefont {Zaccanti}}, \bibinfo {author} {\bibfnamefont
  {M.}~\bibnamefont {Inguscio}}, \ and\ \bibinfo {author} {\bibfnamefont
  {G.}~\bibnamefont {Roati}},\ }\bibfield  {title} {\bibinfo {title} {\emph
  {Josephson effect in fermionic superfluids across the BEC-BCS crossover}},\
  }\href {\doibase 10.1126/science.aac9725} {\bibfield  {journal} {\bibinfo
  {journal} {Science}\ }\textbf {\bibinfo {volume} {350}},\ \bibinfo {pages}
  {1505} (\bibinfo {year} {2015})}\BibitemShut {NoStop}%
\bibitem [{\citenamefont {Husmann}\ \emph {et~al.}(2015)\citenamefont
  {Husmann}, \citenamefont {Uchino}, \citenamefont {Krinner}, \citenamefont
  {Lebrat}, \citenamefont {Giamarchi}, \citenamefont {Esslinger},\ and\
  \citenamefont {Brantut}}]{Husmann2015}%
  \BibitemOpen
  \bibfield  {author} {\bibinfo {author} {\bibfnamefont {D.}~\bibnamefont
  {Husmann}}, \bibinfo {author} {\bibfnamefont {S.}~\bibnamefont {Uchino}},
  \bibinfo {author} {\bibfnamefont {S.}~\bibnamefont {Krinner}}, \bibinfo
  {author} {\bibfnamefont {M.}~\bibnamefont {Lebrat}}, \bibinfo {author}
  {\bibfnamefont {T.}~\bibnamefont {Giamarchi}}, \bibinfo {author}
  {\bibfnamefont {T.}~\bibnamefont {Esslinger}}, \ and\ \bibinfo {author}
  {\bibfnamefont {J.-P.}\ \bibnamefont {Brantut}},\ }\bibfield  {title}
  {\bibinfo {title} {\emph {Connecting strongly correlated superfluids by a
  quantum point contact}},\ }\href {\doibase 10.1126/science.aac9584}
  {\bibfield  {journal} {\bibinfo  {journal} {Science}\ }\textbf {\bibinfo
  {volume} {350}},\ \bibinfo {pages} {1498} (\bibinfo {year}
  {2015})}\BibitemShut {NoStop}%
\bibitem [{\citenamefont {Burchianti}\ \emph {et~al.}(2018)\citenamefont
  {Burchianti}, \citenamefont {Scazza}, \citenamefont {Amico}, \citenamefont
  {Valtolina}, \citenamefont {Seman}, \citenamefont {Fort}, \citenamefont
  {Zaccanti}, \citenamefont {Inguscio},\ and\ \citenamefont
  {Roati}}]{Burchianti2018}%
  \BibitemOpen
  \bibfield  {author} {\bibinfo {author} {\bibfnamefont {A.}~\bibnamefont
  {Burchianti}}, \bibinfo {author} {\bibfnamefont {F.}~\bibnamefont {Scazza}},
  \bibinfo {author} {\bibfnamefont {A.}~\bibnamefont {Amico}}, \bibinfo
  {author} {\bibfnamefont {G.}~\bibnamefont {Valtolina}}, \bibinfo {author}
  {\bibfnamefont {J.~A.}\ \bibnamefont {Seman}}, \bibinfo {author}
  {\bibfnamefont {C.}~\bibnamefont {Fort}}, \bibinfo {author} {\bibfnamefont
  {M.}~\bibnamefont {Zaccanti}}, \bibinfo {author} {\bibfnamefont
  {M.}~\bibnamefont {Inguscio}}, \ and\ \bibinfo {author} {\bibfnamefont
  {G.}~\bibnamefont {Roati}},\ }\bibfield  {title} {\bibinfo {title} {\emph
  {Connecting Dissipation and Phase Slips in a Josephson Junction between
  Fermionic Superfluids}},\ }\href {\doibase 10.1103/PhysRevLett.120.025302}
  {\bibfield  {journal} {\bibinfo  {journal} {Phys. Rev. Lett.}\ }\textbf
  {\bibinfo {volume} {120}},\ \bibinfo {pages} {025302} (\bibinfo {year}
  {2018})}\BibitemShut {NoStop}%
\bibitem [{\citenamefont {Husmann}\ \emph {et~al.}(2018)\citenamefont
  {Husmann}, \citenamefont {Lebrat}, \citenamefont {H{\"a}usler}, \citenamefont
  {Brantut}, \citenamefont {Corman},\ and\ \citenamefont
  {Esslinger}}]{Husmann2018}%
  \BibitemOpen
  \bibfield  {author} {\bibinfo {author} {\bibfnamefont {D.}~\bibnamefont
  {Husmann}}, \bibinfo {author} {\bibfnamefont {M.}~\bibnamefont {Lebrat}},
  \bibinfo {author} {\bibfnamefont {S.}~\bibnamefont {H{\"a}usler}}, \bibinfo
  {author} {\bibfnamefont {J.-P.}\ \bibnamefont {Brantut}}, \bibinfo {author}
  {\bibfnamefont {L.}~\bibnamefont {Corman}}, \ and\ \bibinfo {author}
  {\bibfnamefont {T.}~\bibnamefont {Esslinger}},\ }\bibfield  {title} {\bibinfo
  {title} {\emph {Breakdown of the Wiedemann{\textendash}Franz law in a unitary
  Fermi gas}},\ }\href {\doibase 10.1073/pnas.1803336115} {\bibfield  {journal}
  {\bibinfo  {journal} {Proc. Nat. Acad. Sci.}\ }\textbf {\bibinfo {volume}
  {115}},\ \bibinfo {pages} {8563} (\bibinfo {year} {2018})}\BibitemShut
  {NoStop}%
\bibitem [{\citenamefont {Kwon}\ \emph {et~al.}(2020)\citenamefont {Kwon},
  \citenamefont {Del~Pace}, \citenamefont {Panza}, \citenamefont {Inguscio},
  \citenamefont {Zwerger}, \citenamefont {Zaccanti}, \citenamefont {Scazza},\
  and\ \citenamefont {Roati}}]{Kwon2020}%
  \BibitemOpen
  \bibfield  {author} {\bibinfo {author} {\bibfnamefont {W.~J.}\ \bibnamefont
  {Kwon}}, \bibinfo {author} {\bibfnamefont {G.}~\bibnamefont {Del~Pace}},
  \bibinfo {author} {\bibfnamefont {R.}~\bibnamefont {Panza}}, \bibinfo
  {author} {\bibfnamefont {M.}~\bibnamefont {Inguscio}}, \bibinfo {author}
  {\bibfnamefont {W.}~\bibnamefont {Zwerger}}, \bibinfo {author} {\bibfnamefont
  {M.}~\bibnamefont {Zaccanti}}, \bibinfo {author} {\bibfnamefont
  {F.}~\bibnamefont {Scazza}}, \ and\ \bibinfo {author} {\bibfnamefont
  {G.}~\bibnamefont {Roati}},\ }\bibfield  {title} {\bibinfo {title} {\emph
  {Strongly correlated superfluid order parameters from dc Josephson
  supercurrents}},\ }\href {\doibase 10.1126/science.aaz2463} {\bibfield
  {journal} {\bibinfo  {journal} {Science}\ }\textbf {\bibinfo {volume}
  {369}},\ \bibinfo {pages} {84} (\bibinfo {year} {2020})}\BibitemShut
  {NoStop}%
\bibitem [{\citenamefont {Luick}\ \emph {et~al.}(2020)\citenamefont {Luick},
  \citenamefont {Sobirey}, \citenamefont {Bohlen}, \citenamefont {Singh},
  \citenamefont {Mathey}, \citenamefont {Lompe},\ and\ \citenamefont
  {Moritz}}]{Luick2020}%
  \BibitemOpen
  \bibfield  {author} {\bibinfo {author} {\bibfnamefont {N.}~\bibnamefont
  {Luick}}, \bibinfo {author} {\bibfnamefont {L.}~\bibnamefont {Sobirey}},
  \bibinfo {author} {\bibfnamefont {M.}~\bibnamefont {Bohlen}}, \bibinfo
  {author} {\bibfnamefont {V.~P.}\ \bibnamefont {Singh}}, \bibinfo {author}
  {\bibfnamefont {L.}~\bibnamefont {Mathey}}, \bibinfo {author} {\bibfnamefont
  {T.}~\bibnamefont {Lompe}}, \ and\ \bibinfo {author} {\bibfnamefont
  {H.}~\bibnamefont {Moritz}},\ }\bibfield  {title} {\bibinfo {title} {\emph
  {An ideal Josephson junction in an ultracold two-dimensional Fermi gas}},\
  }\href {\doibase 10.1126/science.aaz2342} {\bibfield  {journal} {\bibinfo
  {journal} {Science}\ }\textbf {\bibinfo {volume} {369}},\ \bibinfo {pages}
  {89} (\bibinfo {year} {2020})}\BibitemShut {NoStop}%
\bibitem [{\citenamefont {Jank\'o}\ \emph {et~al.}(1999)\citenamefont
  {Jank\'o}, \citenamefont {Kosztin}, \citenamefont {Levin}, \citenamefont
  {Norman},\ and\ \citenamefont {Scalapino}}]{Janko1999}%
  \BibitemOpen
  \bibfield  {author} {\bibinfo {author} {\bibfnamefont {B.}~\bibnamefont
  {Jank\'o}}, \bibinfo {author} {\bibfnamefont {I.}~\bibnamefont {Kosztin}},
  \bibinfo {author} {\bibfnamefont {K.}~\bibnamefont {Levin}}, \bibinfo
  {author} {\bibfnamefont {M.~R.}\ \bibnamefont {Norman}}, \ and\ \bibinfo
  {author} {\bibfnamefont {D.~J.}\ \bibnamefont {Scalapino}},\ }\bibfield
  {title} {\bibinfo {title} {\emph {Incoherent Pair Tunneling as a Probe of the
  Cuprate Pseudogap}},\ }\href {\doibase 10.1103/PhysRevLett.82.4304}
  {\bibfield  {journal} {\bibinfo  {journal} {Phys. Rev. Lett.}\ }\textbf
  {\bibinfo {volume} {82}},\ \bibinfo {pages} {4304} (\bibinfo {year}
  {1999})}\BibitemShut {NoStop}%
\bibitem [{\citenamefont {Bergeal}\ \emph {et~al.}(2008)\citenamefont
  {Bergeal}, \citenamefont {Lesueur}, \citenamefont {Aprili}, \citenamefont
  {Faini}, \citenamefont {Contour},\ and\ \citenamefont
  {Leridon}}]{Bergeal2008}%
  \BibitemOpen
  \bibfield  {author} {\bibinfo {author} {\bibfnamefont {N.}~\bibnamefont
  {Bergeal}}, \bibinfo {author} {\bibfnamefont {J.}~\bibnamefont {Lesueur}},
  \bibinfo {author} {\bibfnamefont {M.}~\bibnamefont {Aprili}}, \bibinfo
  {author} {\bibfnamefont {G.}~\bibnamefont {Faini}}, \bibinfo {author}
  {\bibfnamefont {J.~P.}\ \bibnamefont {Contour}}, \ and\ \bibinfo {author}
  {\bibfnamefont {B.}~\bibnamefont {Leridon}},\ }\bibfield  {title} {\bibinfo
  {title} {\emph {Pairing fluctuations in the pseudogap state of copper-oxide
  superconductors probed by the Josephson effect}},\ }\href
  {https://doi.org/10.1038/nphys1017} {\bibfield  {journal} {\bibinfo
  {journal} {Nature Phys.}\ }\textbf {\bibinfo {volume} {4}},\ \bibinfo {pages}
  {608} (\bibinfo {year} {2008})}\BibitemShut {NoStop}%
\bibitem [{\citenamefont {Pini}\ \emph {et~al.}(2020)\citenamefont {Pini},
  \citenamefont {Pieri}, \citenamefont {Jaeger}, \citenamefont {Denschlag},\
  and\ \citenamefont {Strinati}}]{Pini2019}%
  \BibitemOpen
  \bibfield  {author} {\bibinfo {author} {\bibfnamefont {M.}~\bibnamefont
  {Pini}}, \bibinfo {author} {\bibfnamefont {P.}~\bibnamefont {Pieri}},
  \bibinfo {author} {\bibfnamefont {M.}~\bibnamefont {Jaeger}}, \bibinfo
  {author} {\bibfnamefont {J.~P.~H.}\ \bibnamefont {Denschlag}}, \ and\
  \bibinfo {author} {\bibfnamefont {G.}~\bibnamefont {Strinati}},\ }\bibfield
  {title} {\bibinfo {title} {\emph {Pair correlations in the normal phase of an
  attractive Fermi gas}},\ }\href {https://doi.org/10.1088/1367-2630/ab9ee3}
  {\bibfield  {journal} {\bibinfo  {journal} {New J. Phys.}\ }\textbf {\bibinfo
  {volume} {22}} (\bibinfo {year} {2020})}\BibitemShut {NoStop}%
\bibitem [{\citenamefont {Haussmann}\ and\ \citenamefont
  {Zwerger}(2008)}]{Haussmann2008}%
  \BibitemOpen
  \bibfield  {author} {\bibinfo {author} {\bibfnamefont {R.}~\bibnamefont
  {Haussmann}}\ and\ \bibinfo {author} {\bibfnamefont {W.}~\bibnamefont
  {Zwerger}},\ }\bibfield  {title} {\bibinfo {title} {\emph {Thermodynamics of
  a trapped unitary Fermi gas}},\ }\href {\doibase 10.1103/PhysRevA.78.063602}
  {\bibfield  {journal} {\bibinfo  {journal} {Phys. Rev. A}\ }\textbf {\bibinfo
  {volume} {78}},\ \bibinfo {pages} {063602} (\bibinfo {year}
  {2008})}\BibitemShut {NoStop}%
\bibitem [{SM()}]{SM}%
  \BibitemOpen
  \href@noop {} {}\bibinfo {note} {See Supplemental Material for details on the experimental methods and the theoretical
  modeling tools.}\BibitemShut {Stop}%
\bibitem [{\citenamefont {Giovanazzi}\ \emph {et~al.}(2000)\citenamefont
  {Giovanazzi}, \citenamefont {Smerzi},\ and\ \citenamefont
  {Fantoni}}]{Giovanazzi2000}%
  \BibitemOpen
  \bibfield  {author} {\bibinfo {author} {\bibfnamefont {S.}~\bibnamefont
  {Giovanazzi}}, \bibinfo {author} {\bibfnamefont {A.}~\bibnamefont {Smerzi}},
  \ and\ \bibinfo {author} {\bibfnamefont {S.}~\bibnamefont {Fantoni}},\
  }\bibfield  {title} {\bibinfo {title} {\emph {Josephson Effects in Dilute
  Bose-Einstein Condensates}},\ }\href {\doibase 10.1103/PhysRevLett.84.4521}
  {\bibfield  {journal} {\bibinfo  {journal} {Phys. Rev. Lett.}\ }\textbf
  {\bibinfo {volume} {84}},\ \bibinfo {pages} {4521} (\bibinfo {year}
  {2000})}\BibitemShut {NoStop}%
\bibitem [{\citenamefont {Levy}\ \emph {et~al.}(2007)\citenamefont {Levy},
  \citenamefont {Lahoud}, \citenamefont {Shomroni},\ and\ \citenamefont
  {Steinhauer}}]{Levy2007}%
  \BibitemOpen
  \bibfield  {author} {\bibinfo {author} {\bibfnamefont {S.}~\bibnamefont
  {Levy}}, \bibinfo {author} {\bibfnamefont {E.}~\bibnamefont {Lahoud}},
  \bibinfo {author} {\bibfnamefont {I.}~\bibnamefont {Shomroni}}, \ and\
  \bibinfo {author} {\bibfnamefont {J.}~\bibnamefont {Steinhauer}},\ }\bibfield
   {title} {\bibinfo {title} {\emph {The a.c. and d.c. Josephson effects in a
  Bose--Einstein condensate}},\ }\href {https://doi.org/10.1038/nature06186}
  {\bibfield  {journal} {\bibinfo  {journal} {Nature}\ }\textbf {\bibinfo
  {volume} {449}},\ \bibinfo {pages} {579} (\bibinfo {year}
  {2007})}\BibitemShut {NoStop}%
\bibitem [{\citenamefont {Anderson}\ and\ \citenamefont
  {Goldman}(1969)}]{Anderson1969}%
  \BibitemOpen
  \bibfield  {author} {\bibinfo {author} {\bibfnamefont {J.~T.}\ \bibnamefont
  {Anderson}}\ and\ \bibinfo {author} {\bibfnamefont {A.~M.}\ \bibnamefont
  {Goldman}},\ }\bibfield  {title} {\bibinfo {title} {\emph {Thermal
  Fluctuations and the Josephson Supercurrent}},\ }\href {\doibase
  10.1103/PhysRevLett.23.128} {\bibfield  {journal} {\bibinfo  {journal} {Phys.
  Rev. Lett.}\ }\textbf {\bibinfo {volume} {23}},\ \bibinfo {pages} {128}
  (\bibinfo {year} {1969})}\BibitemShut {NoStop}%
\bibitem [{\citenamefont {Barone}\ and\ \citenamefont
  {Patern\`o}(1982)}]{BaroneBook}%
  \BibitemOpen
  \bibfield  {author} {\bibinfo {author} {\bibfnamefont {A.}~\bibnamefont
  {Barone}}\ and\ \bibinfo {author} {\bibfnamefont {G.}~\bibnamefont
  {Patern\`o}},\ }\href@noop {} {\emph {\bibinfo {title} {Physics and
  Applications of the Josephson Effect}}}\ (\bibinfo  {publisher} {John
  Wiley},\ \bibinfo {address} {New York},\ \bibinfo {year} {1982})\BibitemShut
  {NoStop}%
\bibitem [{\citenamefont {Bloch}(1970)}]{Bloch1970}%
  \BibitemOpen
  \bibfield  {author} {\bibinfo {author} {\bibfnamefont {F.}~\bibnamefont
  {Bloch}},\ }\bibfield  {title} {\bibinfo {title} {\emph {Josephson Effect in
  a Superconducting Ring}},\ }\href {\doibase 10.1103/PhysRevB.2.109}
  {\bibfield  {journal} {\bibinfo  {journal} {Phys. Rev. B}\ }\textbf {\bibinfo
  {volume} {2}},\ \bibinfo {pages} {109} (\bibinfo {year} {1970})}\BibitemShut
  {NoStop}%
\bibitem [{\citenamefont {Ku}\ \emph {et~al.}(2012)\citenamefont {Ku},
  \citenamefont {Sommer}, \citenamefont {Cheuk},\ and\ \citenamefont
  {Zwierlein}}]{Ku2012}%
  \BibitemOpen
  \bibfield  {author} {\bibinfo {author} {\bibfnamefont {M.~J.~H.}\
  \bibnamefont {Ku}}, \bibinfo {author} {\bibfnamefont {A.~T.}\ \bibnamefont
  {Sommer}}, \bibinfo {author} {\bibfnamefont {L.~W.}\ \bibnamefont {Cheuk}}, \
  and\ \bibinfo {author} {\bibfnamefont {M.~W.}\ \bibnamefont {Zwierlein}},\
  }\bibfield  {title} {\bibinfo {title} {\emph {Revealing the Superfluid Lambda
  Transition in the Universal Thermodynamics of a Unitary Fermi Gas}},\ }\href
  {\doibase 10.1126/science.1214987} {\bibfield  {journal} {\bibinfo  {journal}
  {Science}\ }\textbf {\bibinfo {volume} {335}},\ \bibinfo {pages} {563}
  (\bibinfo {year} {2012})}\BibitemShut {NoStop}%
\bibitem [{\citenamefont {Ambegaokar}\ and\ \citenamefont
  {Baratoff}(1963)}]{Ambegaokar1963b}%
  \BibitemOpen
  \bibfield  {author} {\bibinfo {author} {\bibfnamefont {V.}~\bibnamefont
  {Ambegaokar}}\ and\ \bibinfo {author} {\bibfnamefont {A.}~\bibnamefont
  {Baratoff}},\ }\bibfield  {title} {\bibinfo {title} {\emph {Tunneling Between
  Superconductors}},\ }\href {\doibase 10.1103/PhysRevLett.11.104} {\bibfield
  {journal} {\bibinfo  {journal} {Phys. Rev. Lett.}\ }\textbf {\bibinfo
  {volume} {11}},\ \bibinfo {pages} {104} (\bibinfo {year} {1963})}\BibitemShut
  {NoStop}%
\bibitem [{\citenamefont {Spuntarelli}\ \emph {et~al.}(2007)\citenamefont
  {Spuntarelli}, \citenamefont {Pieri},\ and\ \citenamefont
  {Strinati}}]{Spuntarelli2007}%
  \BibitemOpen
  \bibfield  {author} {\bibinfo {author} {\bibfnamefont {A.}~\bibnamefont
  {Spuntarelli}}, \bibinfo {author} {\bibfnamefont {P.}~\bibnamefont {Pieri}},
  \ and\ \bibinfo {author} {\bibfnamefont {G.~C.}\ \bibnamefont {Strinati}},\
  }\bibfield  {title} {\bibinfo {title} {\emph {Josephson Effect throughout the
  BCS-BEC Crossover}},\ }\href {\doibase 10.1103/PhysRevLett.99.040401}
  {\bibfield  {journal} {\bibinfo  {journal} {Phys. Rev. Lett.}\ }\textbf
  {\bibinfo {volume} {99}},\ \bibinfo {pages} {040401} (\bibinfo {year}
  {2007})}\BibitemShut {NoStop}%
\bibitem [{\citenamefont {Zaccanti}\ and\ \citenamefont
  {Zwerger}(2019)}]{Zaccanti2019}%
  \BibitemOpen
  \bibfield  {author} {\bibinfo {author} {\bibfnamefont {M.}~\bibnamefont
  {Zaccanti}}\ and\ \bibinfo {author} {\bibfnamefont {W.}~\bibnamefont
  {Zwerger}},\ }\bibfield  {title} {\bibinfo {title} {\emph {Critical Josephson
  current in BCS-BEC--crossover superfluids}},\ }\href {\doibase
  10.1103/PhysRevA.100.063601} {\bibfield  {journal} {\bibinfo  {journal}
  {Phys. Rev. A}\ }\textbf {\bibinfo {volume} {100}},\ \bibinfo {pages}
  {063601} (\bibinfo {year} {2019})}\BibitemShut {NoStop}%
\bibitem [{\citenamefont {Haussmann}\ \emph {et~al.}(2007)\citenamefont
  {Haussmann}, \citenamefont {Rantner}, \citenamefont {Cerrito},\ and\
  \citenamefont {Zwerger}}]{Haussmann2007}%
  \BibitemOpen
  \bibfield  {author} {\bibinfo {author} {\bibfnamefont {R.}~\bibnamefont
  {Haussmann}}, \bibinfo {author} {\bibfnamefont {W.}~\bibnamefont {Rantner}},
  \bibinfo {author} {\bibfnamefont {S.}~\bibnamefont {Cerrito}}, \ and\
  \bibinfo {author} {\bibfnamefont {W.}~\bibnamefont {Zwerger}},\ }\bibfield
  {title} {\bibinfo {title} {\emph {Thermodynamics of the BCS-BEC crossover}},\
  }\href {\doibase 10.1103/PhysRevA.75.023610} {\bibfield  {journal} {\bibinfo
  {journal} {Phys. Rev. A}\ }\textbf {\bibinfo {volume} {75}},\ \bibinfo
  {pages} {023610} (\bibinfo {year} {2007})}\BibitemShut {NoStop}%
\bibitem [{\citenamefont {Halperin}\ \emph {et~al.}(2010)\citenamefont
  {Halperin}, \citenamefont {Refael},\ and\ \citenamefont
  {Demler}}]{Halperin2011}%
  \BibitemOpen
  \bibfield  {author} {\bibinfo {author} {\bibfnamefont {B.~I.}\ \bibnamefont
  {Halperin}}, \bibinfo {author} {\bibfnamefont {G.}~\bibnamefont {Refael}}, \
  and\ \bibinfo {author} {\bibfnamefont {E.}~\bibnamefont {Demler}},\
  }\bibfield  {title} {\bibinfo {title} {\emph {Resistance In
  Superconductors}},\ }\href {\doibase 10.1142/S021797921005644X} {\bibfield
  {journal} {\bibinfo  {journal} {Int. J. Mod. Phys. B}\ }\textbf {\bibinfo
  {volume} {24}},\ \bibinfo {pages} {4039} (\bibinfo {year}
  {2010})}\BibitemShut {NoStop}%
\bibitem [{\citenamefont {Gor’kov}\ and\ \citenamefont
  {Melik-Barkhudarov}(1961)}]{gor1961}%
  \BibitemOpen
  \bibfield  {author} {\bibinfo {author} {\bibfnamefont {L.}~\bibnamefont
  {Gor’kov}}\ and\ \bibinfo {author} {\bibfnamefont {T.}~\bibnamefont
  {Melik-Barkhudarov}},\ }\bibfield  {title} {\bibinfo {title} {\emph
  {Contribution to the theory of superfluidity in an imperfect Fermi gas}},\
  }\href@noop {} {\bibfield  {journal} {\bibinfo  {journal} {Sov. Phys. JETP}\
  }\textbf {\bibinfo {volume} {13}},\ \bibinfo {pages} {1018} (\bibinfo {year}
  {1961})}\BibitemShut {NoStop}%
\bibitem [{\citenamefont {Uchino}\ and\ \citenamefont
  {Brantut}(2020)}]{Uchino2019}%
  \BibitemOpen
  \bibfield  {author} {\bibinfo {author} {\bibfnamefont {S.}~\bibnamefont
  {Uchino}}\ and\ \bibinfo {author} {\bibfnamefont {J.-P.}\ \bibnamefont
  {Brantut}},\ }\bibfield  {title} {\bibinfo {title} {\emph {Bosonic superfluid
  transport in a quantum point contact}},\ }\href {\doibase
  10.1103/PhysRevResearch.2.023284} {\bibfield  {journal} {\bibinfo  {journal}
  {Phys. Rev. Res.}\ }\textbf {\bibinfo {volume} {2}},\ \bibinfo {pages}
  {023284} (\bibinfo {year} {2020})}\BibitemShut {NoStop}%
\bibitem [{\citenamefont {Haussmann}\ \emph {et~al.}(2009)\citenamefont
  {Haussmann}, \citenamefont {Punk},\ and\ \citenamefont
  {Zwerger}}]{Haussmann2009}%
  \BibitemOpen
  \bibfield  {author} {\bibinfo {author} {\bibfnamefont {R.}~\bibnamefont
  {Haussmann}}, \bibinfo {author} {\bibfnamefont {M.}~\bibnamefont {Punk}}, \
  and\ \bibinfo {author} {\bibfnamefont {W.}~\bibnamefont {Zwerger}},\
  }\bibfield  {title} {\bibinfo {title} {\emph {Spectral functions and rf
  response of ultracold fermionic atoms}},\ }\href {\doibase
  10.1103/PhysRevA.80.063612} {\bibfield  {journal} {\bibinfo  {journal} {Phys.
  Rev. A}\ }\textbf {\bibinfo {volume} {80}},\ \bibinfo {pages} {063612}
  (\bibinfo {year} {2009})}\BibitemShut {NoStop}%
\bibitem [{\citenamefont {Magierski}\ \emph {et~al.}(2009)\citenamefont
  {Magierski}, \citenamefont {Wlaz\l{}owski}, \citenamefont {Bulgac},\ and\
  \citenamefont {Drut}}]{Magierski2009}%
  \BibitemOpen
  \bibfield  {author} {\bibinfo {author} {\bibfnamefont {P.}~\bibnamefont
  {Magierski}}, \bibinfo {author} {\bibfnamefont {G.}~\bibnamefont
  {Wlaz\l{}owski}}, \bibinfo {author} {\bibfnamefont {A.}~\bibnamefont
  {Bulgac}}, \ and\ \bibinfo {author} {\bibfnamefont {J.~E.}\ \bibnamefont
  {Drut}},\ }\bibfield  {title} {\bibinfo {title} {\emph {Finite-Temperature
  Pairing Gap of a Unitary Fermi Gas by Quantum Monte Carlo Calculations}},\
  }\href {\doibase 10.1103/PhysRevLett.103.210403} {\bibfield  {journal}
  {\bibinfo  {journal} {Phys. Rev. Lett.}\ }\textbf {\bibinfo {volume} {103}},\
  \bibinfo {pages} {210403} (\bibinfo {year} {2009})}\BibitemShut {NoStop}%
\bibitem [{\citenamefont {Jensen}\ \emph {et~al.}(2019)\citenamefont {Jensen},
  \citenamefont {Gilbreth},\ and\ \citenamefont {Alhassid}}]{Jensen2019}%
  \BibitemOpen
  \bibfield  {author} {\bibinfo {author} {\bibfnamefont {S.}~\bibnamefont
  {Jensen}}, \bibinfo {author} {\bibfnamefont {C.~N.}\ \bibnamefont
  {Gilbreth}}, \ and\ \bibinfo {author} {\bibfnamefont {Y.}~\bibnamefont
  {Alhassid}},\ }\bibfield  {title} {\bibinfo {title} {\emph {The pseudogap
  regime in the unitary Fermi gas}},\ }\href {\doibase
  10.1140/epjst/e2019-800105-y} {\bibfield  {journal} {\bibinfo  {journal} {EPJ
  ST}\ }\textbf {\bibinfo {volume} {227}},\ \bibinfo {pages} {2241} (\bibinfo
  {year} {2019})}\BibitemShut {NoStop}%
\bibitem [{\citenamefont {Sekino}\ \emph {et~al.}(2020)\citenamefont {Sekino},
  \citenamefont {Tajima},\ and\ \citenamefont {Uchino}}]{Sekino2020}%
  \BibitemOpen
  \bibfield  {author} {\bibinfo {author} {\bibfnamefont {Y.}~\bibnamefont
  {Sekino}}, \bibinfo {author} {\bibfnamefont {H.}~\bibnamefont {Tajima}}, \
  and\ \bibinfo {author} {\bibfnamefont {S.}~\bibnamefont {Uchino}},\
  }\bibfield  {title} {\bibinfo {title} {\emph {Mesoscopic spin transport
  between strongly interacting Fermi gases}},\ }\href {\doibase
  10.1103/PhysRevResearch.2.023152} {\bibfield  {journal} {\bibinfo  {journal}
  {Phys. Rev. Res.}\ }\textbf {\bibinfo {volume} {2}},\ \bibinfo {pages}
  {023152} (\bibinfo {year} {2020})}\BibitemShut {NoStop}%
\bibitem [{\citenamefont {Riedl}\ \emph {et~al.}(2011)\citenamefont {Riedl},
  \citenamefont {Guajardo}, \citenamefont {Kohstall}, \citenamefont
  {Denschlag},\ and\ \citenamefont {Grimm}}]{Riedl2011}%
  \BibitemOpen
  \bibfield  {author} {\bibinfo {author} {\bibfnamefont {S.}~\bibnamefont
  {Riedl}}, \bibinfo {author} {\bibfnamefont {E.~R.~S.}\ \bibnamefont
  {Guajardo}}, \bibinfo {author} {\bibfnamefont {C.}~\bibnamefont {Kohstall}},
  \bibinfo {author} {\bibfnamefont {J.~H.}\ \bibnamefont {Denschlag}}, \ and\
  \bibinfo {author} {\bibfnamefont {R.}~\bibnamefont {Grimm}},\ }\bibfield
  {title} {\bibinfo {title} {\emph {Superfluid quenching of the moment of
  inertia in a strongly interacting Fermi gas}},\ }\href {\doibase
  10.1088/1367-2630/13/3/035003} {\bibfield  {journal} {\bibinfo  {journal}
  {New J. Phys.}\ }\textbf {\bibinfo {volume} {13}},\ \bibinfo {pages} {035003}
  (\bibinfo {year} {2011})}\BibitemShut {NoStop}%
\bibitem [{\citenamefont {Sidorenkov}\ \emph {et~al.}(2013)\citenamefont
  {Sidorenkov}, \citenamefont {Tey}, \citenamefont {Grimm}, \citenamefont
  {Hou}, \citenamefont {Pitaevskii},\ and\ \citenamefont
  {Stringari}}]{Sidorenkov2013}%
  \BibitemOpen
  \bibfield  {author} {\bibinfo {author} {\bibfnamefont {L.~A.}\ \bibnamefont
  {Sidorenkov}}, \bibinfo {author} {\bibfnamefont {M.~K.}\ \bibnamefont {Tey}},
  \bibinfo {author} {\bibfnamefont {R.}~\bibnamefont {Grimm}}, \bibinfo
  {author} {\bibfnamefont {Y.-H.}\ \bibnamefont {Hou}}, \bibinfo {author}
  {\bibfnamefont {L.}~\bibnamefont {Pitaevskii}}, \ and\ \bibinfo {author}
  {\bibfnamefont {S.}~\bibnamefont {Stringari}},\ }\bibfield  {title} {\bibinfo
  {title} {\emph {Second sound and the superfluid fraction in a Fermi gas with
  resonant interactions}},\ }\href {\doibase 10.1038/nature12136} {\bibfield
  {journal} {\bibinfo  {journal} {Nature}\ }\textbf {\bibinfo {volume} {498}},\
  \bibinfo {pages} {78} (\bibinfo {year} {2013})}\BibitemShut {NoStop}%
\bibitem [{\citenamefont {Scalapino}(1970)}]{Scalapino1970}%
  \BibitemOpen
  \bibfield  {author} {\bibinfo {author} {\bibfnamefont {D.~J.}\ \bibnamefont
  {Scalapino}},\ }\bibfield  {title} {\bibinfo {title} {\emph {Pair Tunneling
  as a Probe of Fluctuations in Superconductors}},\ }\href {\doibase
  10.1103/PhysRevLett.24.1052} {\bibfield  {journal} {\bibinfo  {journal}
  {Phys. Rev. Lett.}\ }\textbf {\bibinfo {volume} {24}},\ \bibinfo {pages}
  {1052} (\bibinfo {year} {1970})}\BibitemShut {NoStop}%
\bibitem [{\citenamefont {Anderson}\ and\ \citenamefont
  {Goldman}(1970)}]{Anderson1970}%
  \BibitemOpen
  \bibfield  {author} {\bibinfo {author} {\bibfnamefont {J.~T.}\ \bibnamefont
  {Anderson}}\ and\ \bibinfo {author} {\bibfnamefont {A.~M.}\ \bibnamefont
  {Goldman}},\ }\bibfield  {title} {\bibinfo {title} {\emph {Experimental
  Determination of the Pair Susceptibility of a Superconductor}},\ }\href
  {\doibase 10.1103/PhysRevLett.25.743} {\bibfield  {journal} {\bibinfo
  {journal} {Phys. Rev. Lett.}\ }\textbf {\bibinfo {volume} {25}},\ \bibinfo
  {pages} {743} (\bibinfo {year} {1970})}\BibitemShut {NoStop}%
\end{thebibliography}

\begin{thebibliography}{24}%
\makeatletter
\providecommand \@ifxundefined [1]{%
 \@ifx{#1\undefined}
}%
\providecommand \@ifnum [1]{%
 \ifnum #1\expandafter \@firstoftwo
 \else \expandafter \@secondoftwo
 \fi
}%
\providecommand \@ifx [1]{%
 \ifx #1\expandafter \@firstoftwo
 \else \expandafter \@secondoftwo
 \fi
}%
\providecommand \natexlab [1]{#1}%
\providecommand \enquote  [1]{``#1''}%
\providecommand \bibnamefont  [1]{#1}%
\providecommand \bibfnamefont [1]{#1}%
\providecommand \citenamefont [1]{#1}%
\providecommand \href@noop [0]{\@secondoftwo}%
\providecommand \href [0]{\begingroup \@sanitize@url \@href}%
\providecommand \@href[1]{\@@startlink{#1}\@@href}%
\providecommand \@@href[1]{\endgroup#1\@@endlink}%
\providecommand \@sanitize@url [0]{\catcode `\\12\catcode `\$12\catcode
  `\&12\catcode `\#12\catcode `\^12\catcode `\_12\catcode `\%12\relax}%
\providecommand \@@startlink[1]{}%
\providecommand \@@endlink[0]{}%
\providecommand \url  [0]{\begingroup\@sanitize@url \@url }%
\providecommand \@url [1]{\endgroup\@href {#1}{\urlprefix }}%
\providecommand \urlprefix  [0]{URL }%
\providecommand \Eprint [0]{\href }%
\providecommand \doibase [0]{http://dx.doi.org/}%
\providecommand \selectlanguage [0]{\@gobble}%
\providecommand \bibinfo  [0]{\@secondoftwo}%
\providecommand \bibfield  [0]{\@secondoftwo}%
\providecommand \translation [1]{[#1]}%
\providecommand \BibitemOpen [0]{}%
\providecommand \bibitemStop [0]{}%
\providecommand \bibitemNoStop [0]{.\EOS\space}%
\providecommand \EOS [0]{\spacefactor3000\relax}%
\providecommand \BibitemShut  [1]{\csname bibitem#1\endcsname}%
\let\auto@bib@innerbib\@empty
%</preamble>
\bibitem [{\citenamefont {Valtolina}\ \emph {et~al.}(2015)\citenamefont
  {Valtolina}, \citenamefont {Burchianti}, \citenamefont {Amico}, \citenamefont
  {Neri}, \citenamefont {Xhani}, \citenamefont {Seman}, \citenamefont
  {Trombettoni}, \citenamefont {Smerzi}, \citenamefont {Zaccanti},
  \citenamefont {Inguscio},\ and\ \citenamefont {Roati}}]{Valtolina2015sm}%
  \BibitemOpen
  \bibfield  {author} {\bibinfo {author} {\bibfnamefont {G.}~\bibnamefont
  {Valtolina}}, \bibinfo {author} {\bibfnamefont {A.}~\bibnamefont
  {Burchianti}}, \bibinfo {author} {\bibfnamefont {A.}~\bibnamefont {Amico}},
  \bibinfo {author} {\bibfnamefont {E.}~\bibnamefont {Neri}}, \bibinfo {author}
  {\bibfnamefont {K.}~\bibnamefont {Xhani}}, \bibinfo {author} {\bibfnamefont
  {J.~A.}\ \bibnamefont {Seman}}, \bibinfo {author} {\bibfnamefont
  {A.}~\bibnamefont {Trombettoni}}, \bibinfo {author} {\bibfnamefont
  {A.}~\bibnamefont {Smerzi}}, \bibinfo {author} {\bibfnamefont
  {M.}~\bibnamefont {Zaccanti}}, \bibinfo {author} {\bibfnamefont
  {M.}~\bibnamefont {Inguscio}}, \ and\ \bibinfo {author} {\bibfnamefont
  {G.}~\bibnamefont {Roati}},\ }\bibfield  {title} {\bibinfo {title} {\emph
  {Josephson effect in fermionic superfluids across the BEC-BCS crossover}},\
  }\href {\doibase 10.1126/science.aac9725} {\bibfield  {journal} {\bibinfo
  {journal} {Science}\ }\textbf {\bibinfo {volume} {350}},\ \bibinfo {pages}
  {1505} (\bibinfo {year} {2015})}\BibitemShut {NoStop}%
\bibitem [{\citenamefont {Burchianti}\ \emph {et~al.}(2018)\citenamefont
  {Burchianti}, \citenamefont {Scazza}, \citenamefont {Amico}, \citenamefont
  {Valtolina}, \citenamefont {Seman}, \citenamefont {Fort}, \citenamefont
  {Zaccanti}, \citenamefont {Inguscio},\ and\ \citenamefont
  {Roati}}]{Burchianti2018sm}%
  \BibitemOpen
  \bibfield  {author} {\bibinfo {author} {\bibfnamefont {A.}~\bibnamefont
  {Burchianti}}, \bibinfo {author} {\bibfnamefont {F.}~\bibnamefont {Scazza}},
  \bibinfo {author} {\bibfnamefont {A.}~\bibnamefont {Amico}}, \bibinfo
  {author} {\bibfnamefont {G.}~\bibnamefont {Valtolina}}, \bibinfo {author}
  {\bibfnamefont {J.~A.}\ \bibnamefont {Seman}}, \bibinfo {author}
  {\bibfnamefont {C.}~\bibnamefont {Fort}}, \bibinfo {author} {\bibfnamefont
  {M.}~\bibnamefont {Zaccanti}}, \bibinfo {author} {\bibfnamefont
  {M.}~\bibnamefont {Inguscio}}, \ and\ \bibinfo {author} {\bibfnamefont
  {G.}~\bibnamefont {Roati}},\ }\bibfield  {title} {\bibinfo {title} {\emph
  {Connecting Dissipation and Phase Slips in a Josephson Junction between
  Fermionic Superfluids}},\ }\href {\doibase 10.1103/PhysRevLett.120.025302}
  {\bibfield  {journal} {\bibinfo  {journal} {Phys. Rev. Lett.}\ }\textbf
  {\bibinfo {volume} {120}},\ \bibinfo {pages} {025302} (\bibinfo {year}
  {2018})}\BibitemShut {NoStop}%
\bibitem [{\citenamefont {Kwon}\ \emph {et~al.}(2020)\citenamefont {Kwon},
  \citenamefont {Del~Pace}, \citenamefont {Panza}, \citenamefont {Inguscio},
  \citenamefont {Zwerger}, \citenamefont {Zaccanti}, \citenamefont {Scazza},\
  and\ \citenamefont {Roati}}]{Kwon2019}%
  \BibitemOpen
  \bibfield  {author} {\bibinfo {author} {\bibfnamefont {W.~J.}\ \bibnamefont
  {Kwon}}, \bibinfo {author} {\bibfnamefont {G.}~\bibnamefont {Del~Pace}},
  \bibinfo {author} {\bibfnamefont {R.}~\bibnamefont {Panza}}, \bibinfo
  {author} {\bibfnamefont {M.}~\bibnamefont {Inguscio}}, \bibinfo {author}
  {\bibfnamefont {W.}~\bibnamefont {Zwerger}}, \bibinfo {author} {\bibfnamefont
  {M.}~\bibnamefont {Zaccanti}}, \bibinfo {author} {\bibfnamefont
  {F.}~\bibnamefont {Scazza}}, \ and\ \bibinfo {author} {\bibfnamefont
  {G.}~\bibnamefont {Roati}},\ }\bibfield  {title} {\bibinfo {title} {\emph
  {Strongly correlated superfluid order parameters from dc Josephson
  supercurrents}},\ }\href {\doibase 10.1126/science.aaz2463} {\bibfield
  {journal} {\bibinfo  {journal} {Science}\ }\textbf {\bibinfo {volume}
  {369}},\ \bibinfo {pages} {84} (\bibinfo {year} {2020})}\BibitemShut
  {NoStop}%
\bibitem [{\citenamefont {Ketterle}\ and\ \citenamefont
  {Zwierlein}(2008)}]{Varenna2008}%
  \BibitemOpen
  \bibfield  {author} {\bibinfo {author} {\bibfnamefont {W.}~\bibnamefont
  {Ketterle}}\ and\ \bibinfo {author} {\bibfnamefont {M.~W.}\ \bibnamefont
  {Zwierlein}},\ }\bibfield  {title} {\bibinfo {title} {\emph {Making, probing
  and understanding ultracold Fermi gases}},\ }in\ \href {\doibase
  10.3254/978-1-58603-846-5-95} {\emph {\bibinfo {booktitle} {Ultra-cold Fermi
  gases}}},\ \bibinfo {series} {Proceedings of the International School of
  Physics ``Enrico Fermi"}, Vol.\ \bibinfo {volume} {164},\ \bibinfo {editor}
  {edited by\ \bibinfo {editor} {\bibfnamefont {M.}~\bibnamefont {Inguscio}},
  \bibinfo {editor} {\bibfnamefont {W.}~\bibnamefont {Ketterle}}, \ and\
  \bibinfo {editor} {\bibfnamefont {C.}~\bibnamefont {Salomon}}}\ (\bibinfo
  {publisher} {IOS press},\ \bibinfo {address} {Amsterdam},\ \bibinfo {year}
  {2008})\ pp.\ \bibinfo {pages} {95--287}\BibitemShut {NoStop}%
\bibitem [{\citenamefont {Ku}\ \emph {et~al.}(2012)\citenamefont {Ku},
  \citenamefont {Sommer}, \citenamefont {Cheuk},\ and\ \citenamefont
  {Zwierlein}}]{Ku2012sm}%
  \BibitemOpen
  \bibfield  {author} {\bibinfo {author} {\bibfnamefont {M.~J.~H.}\
  \bibnamefont {Ku}}, \bibinfo {author} {\bibfnamefont {A.~T.}\ \bibnamefont
  {Sommer}}, \bibinfo {author} {\bibfnamefont {L.~W.}\ \bibnamefont {Cheuk}}, \
  and\ \bibinfo {author} {\bibfnamefont {M.~W.}\ \bibnamefont {Zwierlein}},\
  }\bibfield  {title} {\bibinfo {title} {\emph {Revealing the Superfluid Lambda
  Transition in the Universal Thermodynamics of a Unitary Fermi Gas}},\ }\href
  {\doibase 10.1126/science.1214987} {\bibfield  {journal} {\bibinfo  {journal}
  {Science}\ }\textbf {\bibinfo {volume} {335}},\ \bibinfo {pages} {563}
  (\bibinfo {year} {2012})}\BibitemShut {NoStop}%
\bibitem [{\citenamefont {Guajardo}\ \emph {et~al.}(2013)\citenamefont
  {Guajardo}, \citenamefont {Tey}, \citenamefont {Sidorenkov},\ and\
  \citenamefont {Grimm}}]{guajardo2013}%
  \BibitemOpen
  \bibfield  {author} {\bibinfo {author} {\bibfnamefont {E.~R.~S.}\
  \bibnamefont {Guajardo}}, \bibinfo {author} {\bibfnamefont {M.~K.}\
  \bibnamefont {Tey}}, \bibinfo {author} {\bibfnamefont {L.~A.}\ \bibnamefont
  {Sidorenkov}}, \ and\ \bibinfo {author} {\bibfnamefont {R.}~\bibnamefont
  {Grimm}},\ }\bibfield  {title} {\bibinfo {title} {\emph {Higher-nodal
  collective modes in a resonantly interacting Fermi gas}},\ }\href
  {https://journals.aps.org/pra/abstract/10.1103/PhysRevA.87.063601} {\bibfield
   {journal} {\bibinfo  {journal} {Phys. Rev. A}\ }\textbf {\bibinfo {volume}
  {87}},\ \bibinfo {pages} {063601} (\bibinfo {year} {2013})}\BibitemShut
  {NoStop}%
\bibitem [{\citenamefont {Liu}\ \emph {et~al.}(2009)\citenamefont {Liu},
  \citenamefont {Hu},\ and\ \citenamefont {Drummond}}]{liu2009}%
  \BibitemOpen
  \bibfield  {author} {\bibinfo {author} {\bibfnamefont {X.-J.}\ \bibnamefont
  {Liu}}, \bibinfo {author} {\bibfnamefont {H.}~\bibnamefont {Hu}}, \ and\
  \bibinfo {author} {\bibfnamefont {P.~D.}\ \bibnamefont {Drummond}},\
  }\bibfield  {title} {\bibinfo {title} {\emph {Virial expansion for a strongly
  correlated Fermi gas}},\ }\href@noop {} {\bibfield  {journal} {\bibinfo
  {journal} {Phys. Rev. Lett.}\ }\textbf {\bibinfo {volume} {102}},\ \bibinfo
  {pages} {160401} (\bibinfo {year} {2009})}\BibitemShut {NoStop}%
\bibitem [{\citenamefont {Nascimb{\`e}ne}\ \emph {et~al.}(2010)\citenamefont
  {Nascimb{\`e}ne}, \citenamefont {Navon}, \citenamefont {Jiang}, \citenamefont
  {Chevy},\ and\ \citenamefont {Salomon}}]{Nascimbene2010}%
  \BibitemOpen
  \bibfield  {author} {\bibinfo {author} {\bibfnamefont {S.}~\bibnamefont
  {Nascimb{\`e}ne}}, \bibinfo {author} {\bibfnamefont {N.}~\bibnamefont
  {Navon}}, \bibinfo {author} {\bibfnamefont {K.~J.}\ \bibnamefont {Jiang}},
  \bibinfo {author} {\bibfnamefont {F.}~\bibnamefont {Chevy}}, \ and\ \bibinfo
  {author} {\bibfnamefont {C.}~\bibnamefont {Salomon}},\ }\bibfield  {title}
  {\bibinfo {title} {\emph {Exploring the thermodynamics of a universal Fermi
  gas}},\ }\href {\doibase 10.1038/nature08814} {\bibfield  {journal} {\bibinfo
   {journal} {Nature}\ }\textbf {\bibinfo {volume} {463}},\ \bibinfo {pages}
  {1057} (\bibinfo {year} {2010})}\BibitemShut {NoStop}%
\bibitem [{\citenamefont {Taylor}\ \emph {et~al.}(2009)\citenamefont {Taylor},
  \citenamefont {Hu}, \citenamefont {Liu}, \citenamefont {Pitaevskii},
  \citenamefont {Griffin},\ and\ \citenamefont {Stringari}}]{taylor2009}%
  \BibitemOpen
  \bibfield  {author} {\bibinfo {author} {\bibfnamefont {E.}~\bibnamefont
  {Taylor}}, \bibinfo {author} {\bibfnamefont {H.}~\bibnamefont {Hu}}, \bibinfo
  {author} {\bibfnamefont {X.-J.}\ \bibnamefont {Liu}}, \bibinfo {author}
  {\bibfnamefont {L.}~\bibnamefont {Pitaevskii}}, \bibinfo {author}
  {\bibfnamefont {A.}~\bibnamefont {Griffin}}, \ and\ \bibinfo {author}
  {\bibfnamefont {S.}~\bibnamefont {Stringari}},\ }\bibfield  {title} {\bibinfo
  {title} {\emph {First and second sound in a strongly interacting Fermi
  gas}},\ }\href@noop {} {\bibfield  {journal} {\bibinfo  {journal} {Phys. Rev.
  A}\ }\textbf {\bibinfo {volume} {80}},\ \bibinfo {pages} {053601} (\bibinfo
  {year} {2009})}\BibitemShut {NoStop}%
\bibitem [{\citenamefont {Hou}\ \emph {et~al.}(2013)\citenamefont {Hou},
  \citenamefont {Pitaevskii},\ and\ \citenamefont {Stringari}}]{hou2013}%
  \BibitemOpen
  \bibfield  {author} {\bibinfo {author} {\bibfnamefont {Y.-H.}\ \bibnamefont
  {Hou}}, \bibinfo {author} {\bibfnamefont {L.~P.}\ \bibnamefont {Pitaevskii}},
  \ and\ \bibinfo {author} {\bibfnamefont {S.}~\bibnamefont {Stringari}},\
  }\bibfield  {title} {\bibinfo {title} {\emph {First and second sound in a
  highly elongated Fermi gas at unitarity}},\ }\href@noop {} {\bibfield
  {journal} {\bibinfo  {journal} {Phys. Rev. A}\ }\textbf {\bibinfo {volume}
  {88}},\ \bibinfo {pages} {043630} (\bibinfo {year} {2013})}\BibitemShut
  {NoStop}%
\bibitem [{\citenamefont {Husmann}\ \emph {et~al.}(2015)\citenamefont
  {Husmann}, \citenamefont {Uchino}, \citenamefont {Krinner}, \citenamefont
  {Lebrat}, \citenamefont {Giamarchi}, \citenamefont {Esslinger},\ and\
  \citenamefont {Brantut}}]{husmann2015}%
  \BibitemOpen
  \bibfield  {author} {\bibinfo {author} {\bibfnamefont {D.}~\bibnamefont
  {Husmann}}, \bibinfo {author} {\bibfnamefont {S.}~\bibnamefont {Uchino}},
  \bibinfo {author} {\bibfnamefont {S.}~\bibnamefont {Krinner}}, \bibinfo
  {author} {\bibfnamefont {M.}~\bibnamefont {Lebrat}}, \bibinfo {author}
  {\bibfnamefont {T.}~\bibnamefont {Giamarchi}}, \bibinfo {author}
  {\bibfnamefont {T.}~\bibnamefont {Esslinger}}, \ and\ \bibinfo {author}
  {\bibfnamefont {J.-P.}\ \bibnamefont {Brantut}},\ }\bibfield  {title}
  {\bibinfo {title} {\emph {Connecting strongly correlated superfluids by a
  quantum point contact}},\ }\href@noop {} {\bibfield  {journal} {\bibinfo
  {journal} {Science}\ }\textbf {\bibinfo {volume} {350}},\ \bibinfo {pages}
  {1498} (\bibinfo {year} {2015})}\BibitemShut {NoStop}%
\bibitem [{\citenamefont {Thomas}\ \emph {et~al.}(2005)\citenamefont {Thomas},
  \citenamefont {Kinast},\ and\ \citenamefont {Turlapov}}]{thomas2005}%
  \BibitemOpen
  \bibfield  {author} {\bibinfo {author} {\bibfnamefont {J.~E.}\ \bibnamefont
  {Thomas}}, \bibinfo {author} {\bibfnamefont {J.}~\bibnamefont {Kinast}}, \
  and\ \bibinfo {author} {\bibfnamefont {A.}~\bibnamefont {Turlapov}},\
  }\bibfield  {title} {\bibinfo {title} {\emph {Virial theorem and universality
  in a unitary Fermi gas}},\ }\href@noop {} {\bibfield  {journal} {\bibinfo
  {journal} {Phys. Rev. Lett.}\ }\textbf {\bibinfo {volume} {95}},\ \bibinfo
  {pages} {120402} (\bibinfo {year} {2005})}\BibitemShut {NoStop}%
\bibitem [{\citenamefont {Haussmann}\ \emph {et~al.}(2007)\citenamefont
  {Haussmann}, \citenamefont {Rantner}, \citenamefont {Cerrito},\ and\
  \citenamefont {Zwerger}}]{Haussmann2007sm}%
  \BibitemOpen
  \bibfield  {author} {\bibinfo {author} {\bibfnamefont {R.}~\bibnamefont
  {Haussmann}}, \bibinfo {author} {\bibfnamefont {W.}~\bibnamefont {Rantner}},
  \bibinfo {author} {\bibfnamefont {S.}~\bibnamefont {Cerrito}}, \ and\
  \bibinfo {author} {\bibfnamefont {W.}~\bibnamefont {Zwerger}},\ }\bibfield
  {title} {\bibinfo {title} {\emph {Thermodynamics of the BCS-BEC crossover}},\
  }\href {\doibase 10.1103/PhysRevA.75.023610} {\bibfield  {journal} {\bibinfo
  {journal} {Phys. Rev. A}\ }\textbf {\bibinfo {volume} {75}},\ \bibinfo
  {pages} {023610} (\bibinfo {year} {2007})}\BibitemShut {NoStop}%
\bibitem [{\citenamefont {Carr}\ \emph {et~al.}(2004)\citenamefont {Carr},
  \citenamefont {Shlyapnikov},\ and\ \citenamefont {Castin}}]{Carr2004}%
  \BibitemOpen
  \bibfield  {author} {\bibinfo {author} {\bibfnamefont {L.~D.}\ \bibnamefont
  {Carr}}, \bibinfo {author} {\bibfnamefont {G.~V.}\ \bibnamefont
  {Shlyapnikov}}, \ and\ \bibinfo {author} {\bibfnamefont {Y.}~\bibnamefont
  {Castin}},\ }\bibfield  {title} {\bibinfo {title} {\emph {Achieving a BCS
  Transition in an Atomic Fermi Gas}},\ }\href {\doibase
  10.1103/PhysRevLett.92.150404} {\bibfield  {journal} {\bibinfo  {journal}
  {Phys. Rev. Lett.}\ }\textbf {\bibinfo {volume} {92}},\ \bibinfo {pages}
  {150404} (\bibinfo {year} {2004})}\BibitemShut {NoStop}%
\bibitem [{\citenamefont {Meier}\ and\ \citenamefont
  {Zwerger}(2001)}]{Meier2001sm}%
  \BibitemOpen
  \bibfield  {author} {\bibinfo {author} {\bibfnamefont {F.}~\bibnamefont
  {Meier}}\ and\ \bibinfo {author} {\bibfnamefont {W.}~\bibnamefont
  {Zwerger}},\ }\bibfield  {title} {\bibinfo {title} {\emph {Josephson
  tunneling between weakly interacting Bose-Einstein condensates}},\ }\href
  {\doibase 10.1103/PhysRevA.64.033610} {\bibfield  {journal} {\bibinfo
  {journal} {Phys. Rev. A}\ }\textbf {\bibinfo {volume} {64}},\ \bibinfo
  {pages} {033610} (\bibinfo {year} {2001})}\BibitemShut {NoStop}%
\bibitem [{\citenamefont {Giovanazzi}\ \emph {et~al.}(2000)\citenamefont
  {Giovanazzi}, \citenamefont {Smerzi},\ and\ \citenamefont
  {Fantoni}}]{Giovanazzi2000sm}%
  \BibitemOpen
  \bibfield  {author} {\bibinfo {author} {\bibfnamefont {S.}~\bibnamefont
  {Giovanazzi}}, \bibinfo {author} {\bibfnamefont {A.}~\bibnamefont {Smerzi}},
  \ and\ \bibinfo {author} {\bibfnamefont {S.}~\bibnamefont {Fantoni}},\
  }\bibfield  {title} {\bibinfo {title} {\emph {Josephson Effects in Dilute
  Bose-Einstein Condensates}},\ }\href {\doibase 10.1103/PhysRevLett.84.4521}
  {\bibfield  {journal} {\bibinfo  {journal} {Phys. Rev. Lett.}\ }\textbf
  {\bibinfo {volume} {84}},\ \bibinfo {pages} {4521} (\bibinfo {year}
  {2000})}\BibitemShut {NoStop}%
\bibitem [{\citenamefont {Eckel}\ \emph {et~al.}(2016)\citenamefont {Eckel},
  \citenamefont {Lee}, \citenamefont {Jendrzejewski}, \citenamefont {Lobb},
  \citenamefont {Campbell},\ and\ \citenamefont {Hill}}]{Eckel2016}%
  \BibitemOpen
  \bibfield  {author} {\bibinfo {author} {\bibfnamefont {S.}~\bibnamefont
  {Eckel}}, \bibinfo {author} {\bibfnamefont {J.~G.}\ \bibnamefont {Lee}},
  \bibinfo {author} {\bibfnamefont {F.}~\bibnamefont {Jendrzejewski}}, \bibinfo
  {author} {\bibfnamefont {C.~J.}\ \bibnamefont {Lobb}}, \bibinfo {author}
  {\bibfnamefont {G.~K.}\ \bibnamefont {Campbell}}, \ and\ \bibinfo {author}
  {\bibfnamefont {W.~T.}\ \bibnamefont {Hill}},\ }\bibfield  {title} {\bibinfo
  {title} {\emph {Contact resistance and phase slips in mesoscopic
  superfluid-atom transport}},\ }\href {\doibase 10.1103/PhysRevA.93.063619}
  {\bibfield  {journal} {\bibinfo  {journal} {Phys. Rev. A}\ }\textbf {\bibinfo
  {volume} {93}},\ \bibinfo {pages} {063619} (\bibinfo {year}
  {2016})}\BibitemShut {NoStop}%
\bibitem [{\citenamefont {Zaccanti}\ and\ \citenamefont
  {Zwerger}(2019)}]{Zaccanti2019sm}%
  \BibitemOpen
  \bibfield  {author} {\bibinfo {author} {\bibfnamefont {M.}~\bibnamefont
  {Zaccanti}}\ and\ \bibinfo {author} {\bibfnamefont {W.}~\bibnamefont
  {Zwerger}},\ }\bibfield  {title} {\bibinfo {title} {\emph {Critical Josephson
  current in BCS-BEC--crossover superfluids}},\ }\href {\doibase
  10.1103/PhysRevA.100.063601} {\bibfield  {journal} {\bibinfo  {journal}
  {Phys. Rev. A}\ }\textbf {\bibinfo {volume} {100}},\ \bibinfo {pages}
  {063601} (\bibinfo {year} {2019})}\BibitemShut {NoStop}%
\bibitem [{\citenamefont {Prange}(1963)}]{Prange1963}%
  \BibitemOpen
  \bibfield  {author} {\bibinfo {author} {\bibfnamefont {R.~E.}\ \bibnamefont
  {Prange}},\ }\bibfield  {title} {\bibinfo {title} {\emph {Tunneling from a
  Many-Particle Point of View}},\ }\href {\doibase 10.1103/PhysRev.131.1083}
  {\bibfield  {journal} {\bibinfo  {journal} {Phys. Rev.}\ }\textbf {\bibinfo
  {volume} {131}},\ \bibinfo {pages} {1083} (\bibinfo {year}
  {1963})}\BibitemShut {NoStop}%
\bibitem [{\citenamefont {Bloch}(1970)}]{Bloch1970sm}%
  \BibitemOpen
  \bibfield  {author} {\bibinfo {author} {\bibfnamefont {F.}~\bibnamefont
  {Bloch}},\ }\bibfield  {title} {\bibinfo {title} {\emph {Josephson Effect in
  a Superconducting Ring}},\ }\href {\doibase 10.1103/PhysRevB.2.109}
  {\bibfield  {journal} {\bibinfo  {journal} {Phys. Rev. B}\ }\textbf {\bibinfo
  {volume} {2}},\ \bibinfo {pages} {109} (\bibinfo {year} {1970})}\BibitemShut
  {NoStop}%
\bibitem [{\citenamefont {ter Haar}(1975)}]{haar}%
  \BibitemOpen
  \bibfield  {author} {\bibinfo {author} {\bibfnamefont {D.}~\bibnamefont {ter
  Haar}},\ }\href@noop {} {\emph {\bibinfo {title} {Problems in Quantum
  Mechanics}}},\ \bibinfo {edition} {3rd}\ ed.\ (\bibinfo  {publisher} {Pion
  Limited},\ \bibinfo {address} {London},\ \bibinfo {year} {1975})\BibitemShut
  {NoStop}%
\bibitem [{\citenamefont {Goldobin}\ \emph {et~al.}(2007)\citenamefont
  {Goldobin}, \citenamefont {Koelle}, \citenamefont {Kleiner},\ and\
  \citenamefont {Buzdin}}]{goldobin}%
  \BibitemOpen
  \bibfield  {author} {\bibinfo {author} {\bibfnamefont {E.}~\bibnamefont
  {Goldobin}}, \bibinfo {author} {\bibfnamefont {D.}~\bibnamefont {Koelle}},
  \bibinfo {author} {\bibfnamefont {R.}~\bibnamefont {Kleiner}}, \ and\
  \bibinfo {author} {\bibfnamefont {A.}~\bibnamefont {Buzdin}},\ }\bibfield
  {title} {\bibinfo {title} {\emph {Josephson junctions with second harmonic in
  the current-phase relation: Properties of $\ensuremath{\varphi}$
  junctions}},\ }\href {\doibase 10.1103/PhysRevB.76.224523} {\bibfield
  {journal} {\bibinfo  {journal} {Phys. Rev. B}\ }\textbf {\bibinfo {volume}
  {76}},\ \bibinfo {pages} {224523} (\bibinfo {year} {2007})}\BibitemShut
  {NoStop}%
\bibitem [{\citenamefont {Duke}(1969)}]{Duke1969b}%
  \BibitemOpen
  \bibfield  {author} {\bibinfo {author} {\bibfnamefont {C.~B.}\ \bibnamefont
  {Duke}},\ }\bibinfo {title} {\emph {Theory of Metal-Barrier-Metal
  Tunneling}},\ in\ \href {\doibase 10.1007/978-1-4684-1752-4{\_}4} {\emph
  {\bibinfo {booktitle} {Tunneling Phenomena in Solids}}},\ \bibinfo {editor}
  {edited by\ \bibinfo {editor} {\bibfnamefont {E.}~\bibnamefont {Burstein}}\
  and\ \bibinfo {editor} {\bibfnamefont {S.}~\bibnamefont {Lundqvist}}}\
  (\bibinfo  {publisher} {Springer, Boston MA},\ \bibinfo {year} {1969})\ pp.\
  \bibinfo {pages} {31--46}\BibitemShut {NoStop}%
\bibitem [{\citenamefont {Uchino}\ and\ \citenamefont
  {Brantut}(2020)}]{Uchino2019sm}%
  \BibitemOpen
  \bibfield  {author} {\bibinfo {author} {\bibfnamefont {S.}~\bibnamefont
  {Uchino}}\ and\ \bibinfo {author} {\bibfnamefont {J.-P.}\ \bibnamefont
  {Brantut}},\ }\bibfield  {title} {\bibinfo {title} {\emph {Bosonic superfluid
  transport in a quantum point contact}},\ }\href {\doibase
  10.1103/PhysRevResearch.2.023284} {\bibfield  {journal} {\bibinfo  {journal}
  {Phys. Rev. Res.}\ }\textbf {\bibinfo {volume} {2}},\ \bibinfo {pages}
  {023284} (\bibinfo {year} {2020})}\BibitemShut {NoStop}%
\end{thebibliography}
\end{document}